\documentclass[aps,prd,onecolumn,superscriptaddress,floats,amsfonts,amsmath,amssymb,linenumbers,12pt]{revtex4}
\usepackage{graphicx}
\usepackage{epstopdf}
\usepackage[active]{srcltx}
\usepackage{bm}
\usepackage[]{latexsym}

\usepackage{color}

\usepackage{caption2}

\usepackage{float}

\usepackage{hyperref}
\hypersetup{pdfauthor={huchiayu},pdftitle={},
            pdfsubject={},pdfcreator={pdfLaTeX with hyperref (\today)}}

\usepackage{lineno}

\begin{document}

\restylefloat{figure}

\title{Probing Neutrino Flavor Transition Mechanism with Ultra High Energy Astrophysical Neutrinos}


\author{Kwang-Chang Lai}
\email{kcl@mail.cgu.edu.tw}
\affiliation{Center for General Education, Chang Gung University, Kwei-Shan, Taoyuan, 333, Taiwan}
\affiliation{Leung Center for Cosmology and Particle Astrophysics (LeCosPA), National Taiwan University, Taipei, 106, Taiwan}

\author{Guey-Lin Lin}
\affiliation{Institute of Physics, National Chiao Tung University, Hsinchu, 300, Taiwan}
\affiliation{Leung Center for Cosmology and Particle Astrophysics (LeCosPA), National Taiwan University, Taipei, 106, Taiwan}

\author{Tsung-Che Liu}
\affiliation{Leung Center for Cosmology and Particle Astrophysics (LeCosPA), National Taiwan University, Taipei, 106, Taiwan}




\begin{abstract}

Observation of ultra-high energy astrophysical neutrinos and identification of their flavors have been proposed for future neutrino telescopes. The flavor ratio of astrophysical neutrinos observed on the Earth depends on both the initial flavor ratio at the source and flavor transitions taking place during propagations of these neutrinos. The flavor transition mechanisms are well-classified with our model-independent parametrization. We find that a new parameter $R\equiv\phi_e/(\phi_\mu+\phi_\tau)$ can probe directly the flavor transition in the framework of our model-independent parametrization, without the assumption of the $\nu_\mu-\nu_\tau$ symmetry. A few flavor transition models are employed to test our parametrization with this new observable. The observational constraints on flavor transition mechanisms by the new observable are discussed through our model-independent parametrization.

\vspace{3mm}


\end{abstract}

\maketitle

\section{Introduction}

The developments in neutrino telescopes \cite{Berghaus:2008bk,KM3NeT,Gorham:2008yk,Abraham:2009uy,Allison:2009rz} have stimulated ideas of using astrophysical neutrinos as the beam source for probing neutrino flavor transitions \cite{Beacom:2002vi,Barenboim:2003jm,Beacom:2003nh,Beacom:2003zg,Beacom:2003eu,Pakvasa:2004hu,Costantini:2004ap,Bhattacharjee:2005nh,Serpico:2005sz,Serpico:2005bs,Xing:2006uk,Winter:2006ce,Xing:2006xd,Majumdar:2006px,Rodejohann:2006qq,Meloni:2006gv,Blum:2007ie,Hwang:2007na,Pakvasa:2007dc,Choubey:2008di,Maltoni:2008jr,Esmaili:2009dz}. Recently, we have proposed a model independent parametrization for flavor transition mechanisms of astrophysical neutrinos, which propagate a vast distance from the source to the Earth \cite{Lai:2010tj}. Such a parametrization, referred to as the $Q$ matrix parametrization, is physical motivated, and it is a very convenient basis for classifying flavor transition models. We have argued that only $Q_{31}$ and $Q_{33}$ are nonvanishing by assuming the conservation of total neutrino flux and the validity of $\nu_\mu-\nu_\tau$ symmetry \cite{Balantekin:1999dx,Harrison:2002et}. These two nonvanishing matrix elements can be probed by measuring the flavor ratio of astrophysical neutrinos reaching to the Earth. 

Detailed discussions on flavor-ratio measurements in IceCube were first presented in Ref. \cite{Beacom:2003nh}. As neutrinos interact with matters to produce observable signals, the major channel is the changed-current (CC) interaction. The electron produced through $\nu_e$ CC interaction has a large interaction cross section with the medium and produces a shower within a short distance from its production point. Contrary to the electron, the muon produced through $\nu_\mu$ CC interaction can travel a long distance in the medium before it loses all its energy or decays. However, a muon does emit dim light along its propagation so that only those detectors near to the muon track can be triggered. As for $\nu_\tau$ detection, the $\nu_\tau$-induced tau leptons behave differently at different energies for a fixed detector design. For a neutrino telescope such as IceCube\cite{Berghaus:2008bk}, the observable energy range for the double bang event is $3.3~{\rm PeV}<E_\nu<33~{\rm PeV}$. At lower energies, the observation for $\nu_\tau$ event would appear as a shower event since the two bangs cannot be resolved. For an undersea experiment, such as KM3NeT~\cite{KM3NeT}, the observable energy range for the double bang event is similar. The ratio $R_{\rm IceCube}\equiv\phi_\mu/(\phi_e+\phi_\tau)$ can be determined by measuring the muon track to shower ratio\cite{Beacom:2003nh,Winter:2006ce}, which is an appropriate observable for underground/undersea neutrino telescopes. The $\nu_e$ fraction can be extracted from the measurement of this ratio by assuming flavor independence of the neutrino spectrum and the equality of $\nu_\mu$ and $\nu_\tau$ fluxes on the Earth due to the approximate $\nu_\mu-\nu_\tau$ symmetry. Given such a capability in IceCube, we determined the allowed ranges for $Q_{31}$ and $Q_{33}$ as presented in Ref. \cite{Lai:2010tj}. 

As the energy of neutrinos goes higher than a few tens of PeV, the tau lepton range becomes long enough so that a tau lepton can pass through the detector without decay but losing its energy like a muon does. In this case, the signal for $\nu_\tau$ appears like a track event \cite{Liu:2010ae}. This leads to the redefinition of $R$ as $R\equiv\phi_e/(\phi_\mu+\phi_\tau)$, which can be determined by measuring the ratio of shower events induced by electron neutrinos to track events by muon and tau neutrinos. The $\nu_\mu-\nu_\tau$ symmetry is then not a necessity to extract the $\nu_e$ composition. The capability to measure this shower to track ratio in ARA \cite{Allison:2009rz} has been discussed \cite{Lai:2013kja}. With this ratio measured, one can infer the $\nu_e$ composition of the astrophysical high energy neutrino flux. The discrimination between $\nu_e$ and $\nu_\mu+\nu_\tau$ makes possible the determination of the third row of the Q matrix and hence the flavor transition mechanism can be probed. 

In this article, we generalize our previous study to the case without $\nu_\mu-\nu_\tau$ symmetry. This paper is organized as follows. In sec. II, we review the $Q$ representation and present how $R$ and the third row of the $Q$ matrix is related and, hence, the flavor transition mechanism is determined. In Sec. III, we address the flavor discrimination between $\nu_e$ and $\nu_\mu+\nu_\tau$, arguing that the $\phi_e$ composition can be extracted without further distinguishing between $\nu_\mu$ and $\nu_\tau$ in the neutrino telescope. This capability makes $R$ a practical and appropriate observable. Employing the standard three-flavor oscillation and neutrino decay models as examples, we present their $Q$ matrices and values of $R$ in Sec. IV. These results shall pave the way for later discussions on discriminating different flavor transition models with our model independent parametrization. In Sec. V, we study allowed ranges for the element of $Q_{31}$, $Q_{32}$ and $Q_{33}$ for example models. We conclude in Sec. VI.

\section{Review on $Q$ Matrix Formalism}

One can write neutrino flux at the source as  \cite{Lai:2009ke}
\begin{equation}
\Phi_0=\frac{1}{3}V_1+aV_2+bV_3,
\label{define}
\end{equation}
where $\Phi_0=(\phi_{0,e},~\phi_{0,\mu},~\phi_{0,\tau})^T$ with the normalisation $\phi_{0,e}+\phi_{0,\mu}+\phi_{0,\tau}=1$, $V_1=(1,~1,~1)^T$, $V_2=(0,~-1,~1)^T$ and $V_3=(2,~-1,~-1)^T$. The ranges for the source parameters are given by $-1/3+b\leq a\leq1/3-b$ and $-1/6\leq b\leq1/3$. For sources with negligible $\nu_\tau$ fraction, $a=-1/3+b$. The neutrino flux reaching the Earth is then given by
\begin{equation}
\Phi=\kappa V_1+\rho V_2+ \lambda V_3,
\end{equation}
such that \cite{Lai:2010tj}
\begin{equation}
 \left(
 \begin{tabular}{c}
  $\kappa$  \\
  $\rho$  \\
  $\lambda$ 
 \end{tabular}
 \right)
 =
 \left(
 \begin{tabular}{ccc}
  $Q_{11}$ & $Q_{12}$ & $Q_{13}$ \\
  $Q_{21}$ & $Q_{22}$ & $Q_{23}$ \\
  $Q_{31}$ & $Q_{32}$ & $Q_{33}$
 \end{tabular} 
 \right)
  \left(
 \begin{tabular}{c}
  1/3 \\
  a  \\
  b 
 \end{tabular}
 \right),
\end{equation}
where $Q=A^{-1}PA$ with
\begin{equation}
A=
\left(
 \begin{tabular}{ccc}
  1 & 0 & 2 \\
  1 & -1 & -1 \\
  1 & 1 & -1
 \end{tabular} 
 \right).
\end{equation}
In other words, $Q$ is related to $P$ by a similarity transformation where columns of the transformation matrix $A$ correspond to vectors $V_1$, $V_2$ and $V_3$, respectively. The parameters $\kappa$, $\rho$ and $\lambda$ are related to the flux of each neutrino flavor by
\begin{equation}
\phi_e=\kappa+2\lambda, ~\phi_\mu=\kappa-\rho-\lambda, ~\phi_\tau=\kappa+\rho-\lambda, \label{flux}
\end{equation}
with the normalization $\phi_e+\phi_\mu+\phi_\tau=3\kappa$. Since we have chosen the normalization $\phi_{0,e}+\phi_{0,\mu}+\phi_{0,\tau}=1$ for the neutrino flux at the source, the conservation of total neutrino flux during propgations corresponds to $\kappa=1/3$. In general flavour transition models, $\kappa$ could be less than $1/3$ as a consequence of (ordinary) neutrino decaying into invisible states or oscillating into sterile neutrinos. To rewrite Eq. \ref{flux} as
\begin{equation}
\rho=(\phi_\tau-\phi_\mu), \hspace{1cm} \lambda=\phi_e/3-(\phi_\mu+\phi_\tau)/6,
\end{equation}
it is clearly seen that, for a fixed $a$ and $b$, the first row of matrix $Q$ determines the normalization for the total neutrino flux reaching the Earth, the second row of $Q$ determines the breaking of $\nu_\mu-\nu_\tau$ symmetry in the arrival neutrino flux, and the third row of $Q$ determines the flux difference $\phi_e-(\phi_\mu+\phi_\tau)/2$.

For those models which preserve the total neutrino flux, $\sum_{\alpha=e,\mu,\tau}P_{\alpha\beta}=1.$ In the $Q$ matrix representation, these flux-conserving models must give $Q_{11}=1$ and $Q_{12}=Q_{13}=0$. The remaining six matrix elements of $Q$ can be constrained by neutrino telescope measurements. Second, the approximate $\nu_\mu-\nu_\tau$ symmetry makes almost identical the second and third rows of $P$, i.e., $(P_{\mu e},~P_{\mu\mu},~P_{\mu\tau})\approx (P_{\tau e},~P_{\tau\mu},~P_{\tau\tau})$ and also the second and third columns of $P$, i.e., $(P_{e\mu},~P_{\mu\mu},~P_{\tau\mu})^T\approx (P_{e\tau},~P_{\mu\tau},~P_{\tau\tau})^T$. In the $Q$ matrix representation, these properties render $(Q_{21},~Q_{22},~Q_{23})\approx(0,~0,~0)$ and $(Q_{12},~Q_{22},~Q_{32})^T\approx(0,~0,~0)$. In summary, we have seen that the first and second rows of $Q$ as well as the matrix element $Q_{32}$ are already constrained in a simple way by assuming the conservation of total neutrino flux and the validity of approximate $\nu_\mu-\nu_\tau$ symmetry. Hence under these two assumptions, simply the values for $Q_{31}$ and $Q_{33}$ are enough to classify neutrino flavor transition models and the fraction of $\phi_e$ can be extracted from the track to shower ratio. One can therefore probe the transition mechanism with the measurement of the track to shower ratio by neutrino telescopes.






\section{New Observable for Ultra High Energy Neutrino}

It has been demonstrated that \cite{Beacom:2003nh} the event ratio of muon tracks to showers can be used to extract the $\phi_e$ fraction of astrophysical neutrinos, i.e.,  the ratio $R_{\rm IceCube}\equiv\phi_\mu/(\phi_e+\phi_\tau)$, with the approximate $\nu_\mu-\nu_\tau$ symmetry.  
As a result, the relevant elements $Q_{31}$ and $Q_{33}$ can be determined and the flavor transition mechanism is tested~\cite{Lai:2010tj}.

However, the $\nu_\mu-\nu_\tau$ symmetry is broken by the recent confirmation of nonzero $\theta_{13}$ with its value larger than most expectations \cite{DYB,Reno,DChooz}. This symmetry holds because of the maximal mixing $\theta_{23}$ and zero $\theta_{13}$. The nonzero $\theta_{13}$ with relatively large value should benefit the reconstruction of neutrino flavor ratios at astrophysical sources and the discrimination between different astrophysical sources of high energy neutrinos \cite{Lai:2009ke}. However, without the $\nu_\mu-\nu_\tau$ symmetry, we cannot extract the fraction of $\phi_e$ with the measurement 
of track to shower event ratio alone and the flavor transition models cannot be constrained. In order to extract the fraction of $\phi_e$ and then probe the transition mechanism with the $Q$ matrix parametrization with nonzero $\theta_{13}$, we have to further make a difficult measuremnt of  the flux ratio $S_{\rm IceCube}\equiv\phi_e/\phi_\tau$~\cite{Beacom:2003nh}. Fortunately, the situation can be quite different for ultrahigh energy neutrinos.

It has been argued \cite{Liu:2010ae} that, for the astrophysical neutrinos with energies higher than a few tens of PeV, new flux rato should be adopted for the measurements of terrestrial neutrino telescopes. At such high energies, the tau lepton originated from the tau neutrino behaves like a track similar to a muon \cite{Bugaev:2003sw} while the electron neutrino still produces a shower signal. Therefore we are motivated to define more appropriate flux ratio parameters. Meanwhile, radio neutrino telescopes, such as Askaryan Radio Array (ARA), are proposed to observe cosmogenic neutrinos of energy about EeV by detecting Cherenkov radio emissions. These radiations are emitted by showers originated from the ultrahigh energy neutrinos interacting with matter. For cosmogenic neutrinos, the energy of the CC-induced muon or tau lepton is so high that a muon or tau lepton not only emits dim lights but also produce mini-showers along its propagation through the detector fiducial volume. Though the lights emitted from the track can only trigger the nearby optical detectors, a track event can be reconstructed for a muon or tau lepton traversing the detector volume, by detecting the radio emissions from these mini-showers \cite{Lai:2013kja}.  A $\nu_e$ signal of a single, major shower is therefore distinguishable from a $\nu_\mu$ or $\nu_\tau$ track, composed of a sequence of sub-showers. Moreover, the shower induced by $\nu_e$ is an electromagnetic shower while the shower in a $\nu_\mu$ or $\nu_\tau$ track can be either an electromagnetic shower or a hadronic one. These two kinds of showers emit Cherenkov radiations in different patterns. The detection of a hadronic signature can further confirm the existence of a track event.

For $E_\nu>33~{\rm PeV}$, the ratio $R\equiv\phi_e/(\phi_\mu+\phi_\tau)$ is an appropriate and practical parameter for flavor discrimination in water (ice) Chenrenkov and radio-wave neutrino detectors. Although it is difficult to discriminate between $\nu_\mu$ and $\nu_\mu$, the determination of the new parameter $R=\phi_e/(\phi_\mu+\phi_\tau)$ allows one to extract the $\phi_e$ fraction without assuming the $\nu_\mu-\nu_\tau$ symmetry in radio-wave  neutrino telescope experiments.


\section{The Ranges for $Q_{31,32,33}$}



The direct correspondence between the third row of the $Q$ matrix and the flux ratio $R$ defined for ultra high energy neutrinos implies that the measurement of $R$ is already enough to probe the flavor transition mechanism. 
In the case of the conservation of the total neutrino flux, one has $Q_{11}=1$ and $Q_{12}=Q_{13}=0$. In the high energy regime, the remain elements of $Q$ can be constrained through
measuring the ratio $R$. For astrophysical ources with negligible $\nu_{\tau}$ fraction,  we have  $a=-1/3+b$ with parameters $a$ and $b$ defined in Eq.~(\ref{define}). Therefore the ratio $R$
is a function by $b$ such that
\begin{eqnarray}
R(b) & = & -1+\frac{3}{2}[1-(Q_{31}-Q_{32})-3(Q_{32}+Q_{33})b]^{-1}, \nonumber \\ 
         & = & -1+\frac{3}{2}[1-f_{12}-3f_{23}b]^{-1}, \label{RinQ}
\end{eqnarray}
where
\begin{eqnarray}
f_{12} & = & Q_{31}-Q_{32}, \nonumber \\
f_{23} & = & Q_{32}+Q_{33}. \label{fvalue}
\end{eqnarray}
For the pion and muon-damped sources respectively,
\begin{eqnarray}
R_\pi & = & -1+\frac{3}{2}(1-f_{12})^{-1}, \nonumber \\
R_\mu & = & -1+\frac{3}{2}\left(1-f_{12}+\frac{1}{2}f_{23}\right)^{-1}.
\end{eqnarray}
Solving the above equations, we obtain
\begin{eqnarray}
f_{12} & = & 1-\frac{3}{2}(1+R_\pi)^{-1}, \nonumber \\
f_{23} & = & 3(1+R_\mu)^{-1}-3(1+R_\pi)^{-1}. \label{Rtof}
\end{eqnarray}





The parameters $f_{12}$ and $f_{23}$ are good alternatives to $Q_{31}$ and $Q_{33}$ respectively. Flavor transition models can be well classified by $f_{12}$ and $f_{23}$ 
although observations of only two different sources cannot  completely determine the third row of the $Q$ matrix. As noted in \cite{Lai:2010tj}, the $Q$ matrix is related to 
the usual flavor transition matrix $P$ by a similarity transformation. The observation of astrophysical neutrinos can probe the elements of the $Q$ matrix in a 
model independent fashion. Hence, $(f_{12},~f_{23})$  is also a model independent parametrization of flavor transition.

In the following, we illustrate the  $(f_{12},~f_{23})$ parametrization using a few flavor transition models as examples. Discriminating between models with statistical analysis is presented in the next section.

\subsection{Standard neutrino oscillations}

We begin by considering the standard three-flavor neutrino oscillation, in which the probability matrix $P$ for astrophysical neutrinos traversing a vast distance is given by
\begin{equation}
P_{\alpha\beta}\equiv P(\nu_\beta\rightarrow\nu_\alpha)=\sum^3_{i=1}|U_{\beta i}|^2|U_{\alpha i}|^2,
\end{equation}
where $U_{\beta i}$ and $U_{\alpha i}$ are elements of neutrino mixing matrix. The exact form of $P_{\alpha\beta}$ in terms of neutrino mixing parameters is given in the appendix of Ref. \cite{Lai:2009ke}. Since $P_{\alpha\beta}=P_{\beta\alpha}$ in this case, one has $\sum_\alpha P_{\alpha\beta}=\sum_\beta P_{\alpha\beta}=1$. This implies $Q_{21}=Q_{31}=0$ in addition to $Q_{11}=1$ and $Q_{12}=Q_{13}=0$.

For the standard oscillation,
\begin{equation}
Q_{31}=0, \hspace{2mm} Q_{32}=-\frac{1}{2}(P_{e\mu}-P_{e\tau}), \hspace{2mm} Q_{33}=\frac{3}{2}(3P_{ee}-1). \label{Qosc}
\end{equation}
Thus,
\begin{eqnarray}
f_{12} & = & \frac{1}{2}\sum^3_{i=1}|U_{ei}|^2(|U_{\mu i}|^2-|U_{\tau i}|^2), \nonumber \\
f_{23} & = &\sum^3_{i=1}|U_{ei}|^2(|U_{e i}|^2-|U_{\mu i}|^2). \label{stdosc}
\end{eqnarray}
To the first order of $\epsilon=(2\cos2\theta_{23}+\sqrt{2}\sin\theta_{13}\cos\delta)/9$,
\begin{eqnarray}
f_{12} & = & \epsilon\gtrsim0, \nonumber \\
f_{23} & = & \frac{1}{3}-\epsilon\lesssim\frac{1}{3}.
\end{eqnarray}



\subsection{Neutrino Decays}

Flavor transitions of astrophysical neutrinos due to effects of neutrino decays were discussed extensively in Ref. [6]. The simplest case of neutrino decays is that both the heaviest (\textit{H}) and the middle (\textit{M}) mass eigenstates decay into the lightest (\textit{L}) mass eigenstate. If branching ratios of $H\rightarrow L$ and $M\rightarrow L$are both 100\%, the $Q$ matrix is given by [26]
\begin{equation}
Q = \left(
       \begin{tabular}{ccc}
       1 & 0 & 0  \\
       $-3(|U_{\mu j}|^2-|U_{\tau j}|^2)/2$ & 0 & 0  \\
       $|U_{e j}|^2-(|U_{\mu j}|^2+|U_{\tau j}|^2)/2$ & 0 & 0
       \end{tabular}
       \right),
\end{equation}
where $j=1$ for  the normal mass hierarchy and $j=3$ for  the inverted mass hierarchy.

Obviously, $f_{12}\equiv Q_{31}-Q_{32}=Q_{31}=(3|U_{ej}|^2-1)/2$ and $f_{23}\equiv Q_{32}+Q_{33}=0$. To the first order of $D^2\equiv\sin^2\theta_{13}$, the parameters relevant to observations are given by
\begin{eqnarray}
f_{12} & = & \frac{1}{2}-D^2, \nonumber \\
f_{23} & = & 0, \label{dec1norm}
\end{eqnarray}
for $j=1$ and
\begin{eqnarray}
f_{12} & = & -\frac{1}{2}+\frac{3}{2}D^2, \nonumber \\
f_{23} & = & 0, \label{dec1inv}
\end{eqnarray}
for $j=3$. One may consider gene branching ratios for $H\rightarrow L$ and $M\rightarrow L$ decays. However, the resulting neutrino flavor ratio on Earth remains the same in such a scenario.


We next  consider the decay scenario that $H$ decays into both $M$ and $L$ with branching ratios $r$ and $s$ respectively while $M$ does not decay into $L$. Here $r+s=1$ corresponds to the flux conservation case, which we shall adopt in the following discussions on decay models. To the first order of $\epsilon_1=\cos2\theta_{23}-(\sqrt{2}/3)\sin\theta_{13}\cos\delta$ and $\epsilon_2=(1/2)\cos2\theta_{23}-\epsilon_1$, the parameters relevant to observations are then given by
\begin{eqnarray}
f_{12} & = & \frac{1}{6}\big\{(1+s)-\frac{1}{3}[s(\epsilon_1+\epsilon_2)-\epsilon_2]\big\}, \nonumber \\
f_{23} & = & \frac{1}{6}\big\{(1-s)+\frac{1}{3}[s(\epsilon_1+\epsilon_2)-\epsilon_2]\big\}, \label{dec2norm}
\end{eqnarray}
for the normal mass hierarchy and
\begin{eqnarray}
f_{12} & = & \frac{1}{6}\big\{(r-s)+\frac{1}{3}[(1-r+s)\epsilon_1+2\epsilon_2]\big\}, \nonumber \\
f_{23} & = & \frac{1}{6}\big\{2-\frac{1}{3}[(1-r+s)\epsilon_1+2\epsilon_2]\big\}, \label{dec2inv} 
\end{eqnarray}
for the inverted mass hierarchy. Taking into account that $0\leqslant r,~s\leqslant1$, we obtain the ranges for the parameters
\begin{eqnarray}
\frac{1}{6}+\frac{1}{18}\epsilon_2\leqslant & f_{12} & \leqslant\frac{1}{3}-\frac{1}{18}\epsilon_1, \nonumber \\
\frac{1}{18}\epsilon_1\leqslant & f_{23} & \leqslant\frac{1}{6}-\frac{1}{18}\epsilon_2, 
\end{eqnarray}
for the normal mass hierarchy and
\begin{eqnarray}
-\frac{1}{6}+\frac{1}{9}(\epsilon_1+\epsilon_2)\leqslant & f_{12} & \leqslant\frac{1}{6}+\frac{1}{9}\epsilon_2, \nonumber \\
\frac{1}{3}-\frac{1}{9}(\epsilon_1+\epsilon_2)\leqslant & f_{23} & \leqslant\frac{1}{3}-\frac{1}{9}\epsilon_2, 
\end{eqnarray}
for the inverted mass hierarchy. For convenience, let us denote the models described by Eqs. (\ref{dec1norm}) and (\ref{dec1inv}) as $dec1$ scenario and those by Eqs. (\ref{dec2norm}) and (\ref{dec2inv}) as $dec2$ scenario.


\subsection{Quantum Decoherence}

As the last example, we discuss neutrino flavor transitions affected by the decoherence effect from the Planckscale physics \cite{Lisi:2000zt}. In a three-flavor framework, it has been shown that \cite{Gago:2002na,Hooper:2004xr,Anchordoqui:2005gj}
\begin{eqnarray}
P^{\rm dc}_{\alpha\beta} & = & \frac{1}{3}+\big[\frac{1}{2}e^{-\gamma_3d}(U^2_{\beta1}-U^2_{\beta2})(U^2_{\alpha1}-U^2_{\alpha2}) \nonumber \\
                                & + & \frac{1}{6}e^{-\gamma_8d}(U^2_{\beta1}+U^2_{\beta2}-2U^2_{\beta3}) \nonumber  \\
                                & \times & (U^2_{\alpha1}+U^2_{\alpha2}-2U^2_{\alpha3})\big],
\end{eqnarray}
where $\gamma_3$ and $\gamma_8$ are eigenvalues of the decoherence matrix and $d$ is the neutrino propagating distance from the source. The $CP$ phase in the neutrino mixing matrix $U$ has been set to zero. Taking $\gamma_3=\gamma_8=\gamma$, we obtain 
\begin{equation}
Q^{\rm dc}_{11}=1 \hspace{5mm} {\rm and} \hspace{5mm} Q^{\rm dc}_{\alpha\beta}=e^{-\gamma d}Q^{\rm osc}_{\alpha\beta},
\end{equation}
where $Q^{\rm osc}_{\alpha\beta}$ denotes the $Q$ matrix for the standard oscillation in Eq. (\ref{Qosc}). Therefore,
\begin{eqnarray}
f_{12} & = & e^{-\gamma d}\epsilon_0, \nonumber \\
f_{23} & = & e^{-\gamma d}(\frac{1}{3}-\epsilon_0),
\end{eqnarray}
to the first order of $\epsilon_0\equiv\epsilon(\delta=0)=(2\cos2\theta_{23}+\sqrt{2}\sin\theta_{13})/9$. In the absence of the decoherence effect, i.e., $\gamma\rightarrow0$, it is seen that $Q^{\rm dc}$ reduces to the standard oscillation. In the full decoherence case, i.e., $e^{-\gamma d}\rightarrow0$, we have $\kappa=1$ and $\rho=\lambda=0$ such that $\phi_e:\phi_\mu:\phi_\tau=1:1:1$.

\subsection{Pseudo-Dirac Neutrino}

As the last example, let us consider the pseudo-Dirac neutrino scenario\cite{pseudoDirac1,pseudoDirac2,pseudoDirac3}, in which each mass eigenstate of active neutrino is accompanied by a sterile neutrino with degenerated mass. Affected by the existence of sterile states, the neutrino oscillation has been shown to be\cite{Beacom:2003eu,Esmaili:2009dz}
\begin{equation}
P_{\alpha\beta}^{\rm pd}=\sum^3_{i=1}|U_{\beta i}|^2|U_{\alpha i}|^2\cos^2\left[\frac{\Delta m^2_i}{4E_\nu}L(z)\right],
\end{equation}
where $\Delta m^2_i$ is the mass-squared difference between active and sterile states of the $i$-th mass eigenstate and the distance $L(z)$ is given by
\begin{equation}
L(z)=\frac{c}{H_0}\int^z_0\frac{dz'}{(1+z')^2\sqrt{\Omega_m(1+z')^3+\Omega_\Lambda}},
\end{equation}
with $H_0$ the Hubble constant, $\Omega_m$ and $\Omega_\Lambda$, the matter and dark energy densities in units of the critical density, respectively. Taking $\Delta m^2_i=\Delta m^2$ for each $i$, we obtain
\begin{equation}
Q^{\rm pd}_{\alpha\beta}=\cos^2\left[\frac{\Delta m^2}{4E_\nu}L(z)\right]Q_{\alpha\beta},
\end{equation}
In the limit of $\Delta m^2_i=0$ for each $i$, the pseudo-Dirac scenario reduces to the standard oscillation. In the limit of cosmological distances, $L(z)\gg(\Delta m^2_i/4E_\nu)$, the oscillatory phase term will average out such that $Q_{\alpha\beta}^{\rm pd}=(1/2)Q_{\alpha\beta}^{\rm osc}$.

\section{Statistical Analysis}

We have shown that the flavor transitions of astrophysical neutrinos can be classified by the matrix $Q$ and further parametrized by $f_{12}$ and $f_{23}$. In principle, the parameters $f_{12}$ and $f_{23}$ can be determined by measuring the flavor ratios from two different sources, say, a pion source and a muon-damped
source. Practically, the parameters can only be constrained up to a range due to limited accuracies of measurements. Using the measurement of the flux ratio $R$, we perform fitting with
\begin{equation}
\chi^2=\Bigg(\frac{R_{\pi,\rm th}-R_{\pi,\rm exp}}{\sigma_{R_{\pi,\rm exp}}}\Bigg)^2+\Bigg(\frac{R_{\mu,\rm th}-R_{\mu,\rm exp}}{\sigma_{R_{\mu,\rm exp}}}\Bigg)^2, \label{chi2}
\end{equation}
where quantities with the subscript "exp" are experimentally measured flux ratios while quantities with the subscript "th" are theoretically predicted values which depend on parameters defined in Eq. (\ref{fvalue}). Furthermore, $\sigma_{R_{\pi,\rm exp}}=(\Delta R_\pi/R_\pi)R_{\pi,\rm exp}$ and $\sigma_{R_{\mu,\rm exp}}=(\Delta R_\mu/R_\mu)R_{\mu,\rm exp}$ with $\Delta R_\pi$ and $\Delta R_\mu$ the experimental errors in determining $R$ for neutrinos coming from a pion source and muon-damped source, respectively. Should we be able to observe neutrinos from more sources, we have, in general,
\begin{equation}
\chi^2=\sum_i \chi^2_i,
\end{equation}
where
\begin{equation}
\chi^2_i=\Bigg(\frac{R_{i,\rm th}-R_{i,\rm exp}}{\sigma_{R_{i,\rm exp}}}\Bigg)^2,
\end{equation}
with $i$ denoting different sources of astrophysical neutrinos.

\begin{figure}[htbp]
	\includegraphics[width=8cm]{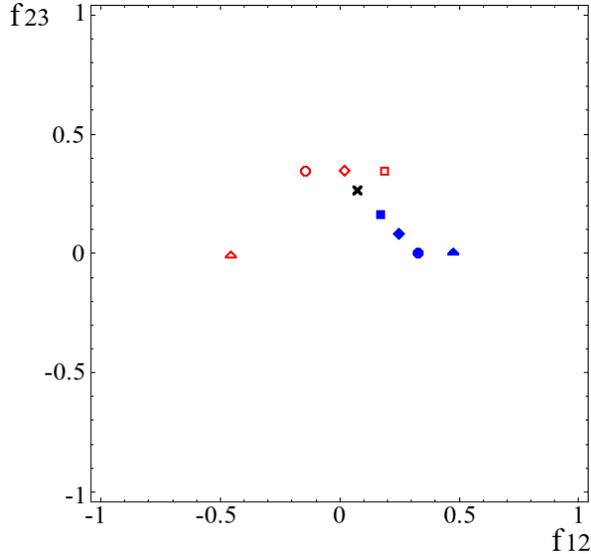}
	\caption{Central values of $f_{12}$ and $f_{23}$ for different flavor transition models. The black cross represents the standard oscillation model and others represent different decay models discussed in the text. The red symbols denote decay models in the normal mass hierarchy and the blue ones denote those in the inverted mass hierarchy. The two triangles represent $dec1$ scenario. The others represent models in $dec2$ scenario and are sampled for branching ratios of $s=0,~0.5,~1$ by squares, diamonds and circles, respectively. The value of $(f_{12},~f_{23})$ for each model is obtained from Eqs. (\ref{stdosc}) and (\ref{dec1norm}-\ref{dec2inv}), respectively.}
	\label{fig:Cvalue}
\end{figure}

We refer to global analysis by \cite{Fogli:2012ua,GonzalezGarcia:2012sz} for neutrino mixing parameters (mixing angles and CP phase). Since the values in both fittings are almost the same, we employ the values from \cite{Fogli:2012ua} in our analysis. In Fig. \ref{fig:Cvalue}, the central values of $f_{12}$ and $f_{23}$ are presented for a few sample models. The cross represents the 
standard neutrino oscillation while the others represent different decay models. The blue and red symbols denote decay models in the normal and inverted mass hierarchies, respectively. The triangles represent models in $dec1$ scenario while the others represent sampled models in $dec2$ scenario. Among these models, the red triangle, described by Eq. (\ref{dec1inv}), is located far from all the others and, thus, anticipated being easily distinguished from other models. One also expects, from this figure, that decay models for different mass hierarchies could be distinguished from each other but the standard oscillation could not be easily distinguished from the $dec2$ scenario in the normal hierarchy.

\begin{figure}[htbp]
	\begin{center}
	\includegraphics[width=8cm]{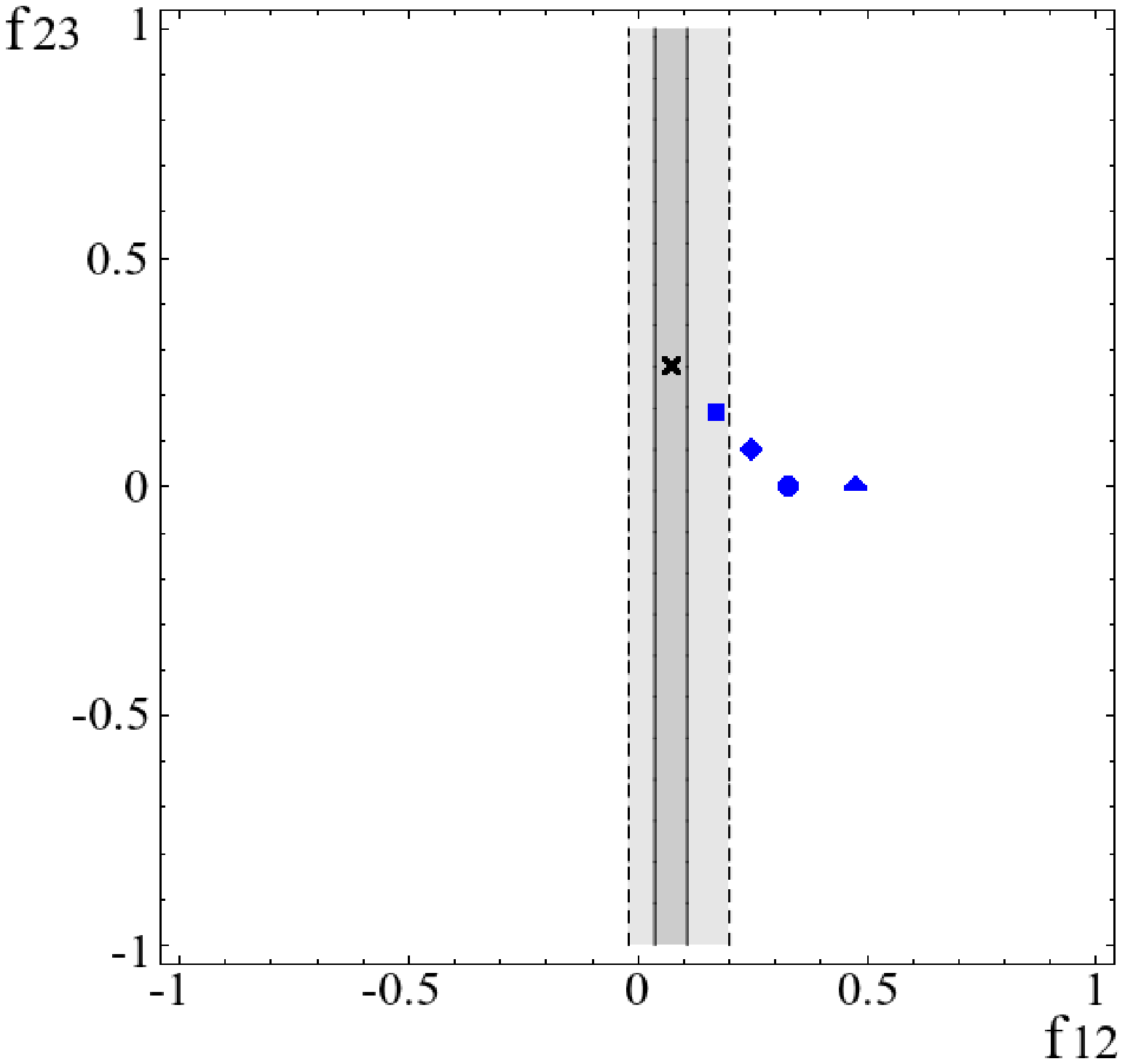}
	\includegraphics[width=8cm]{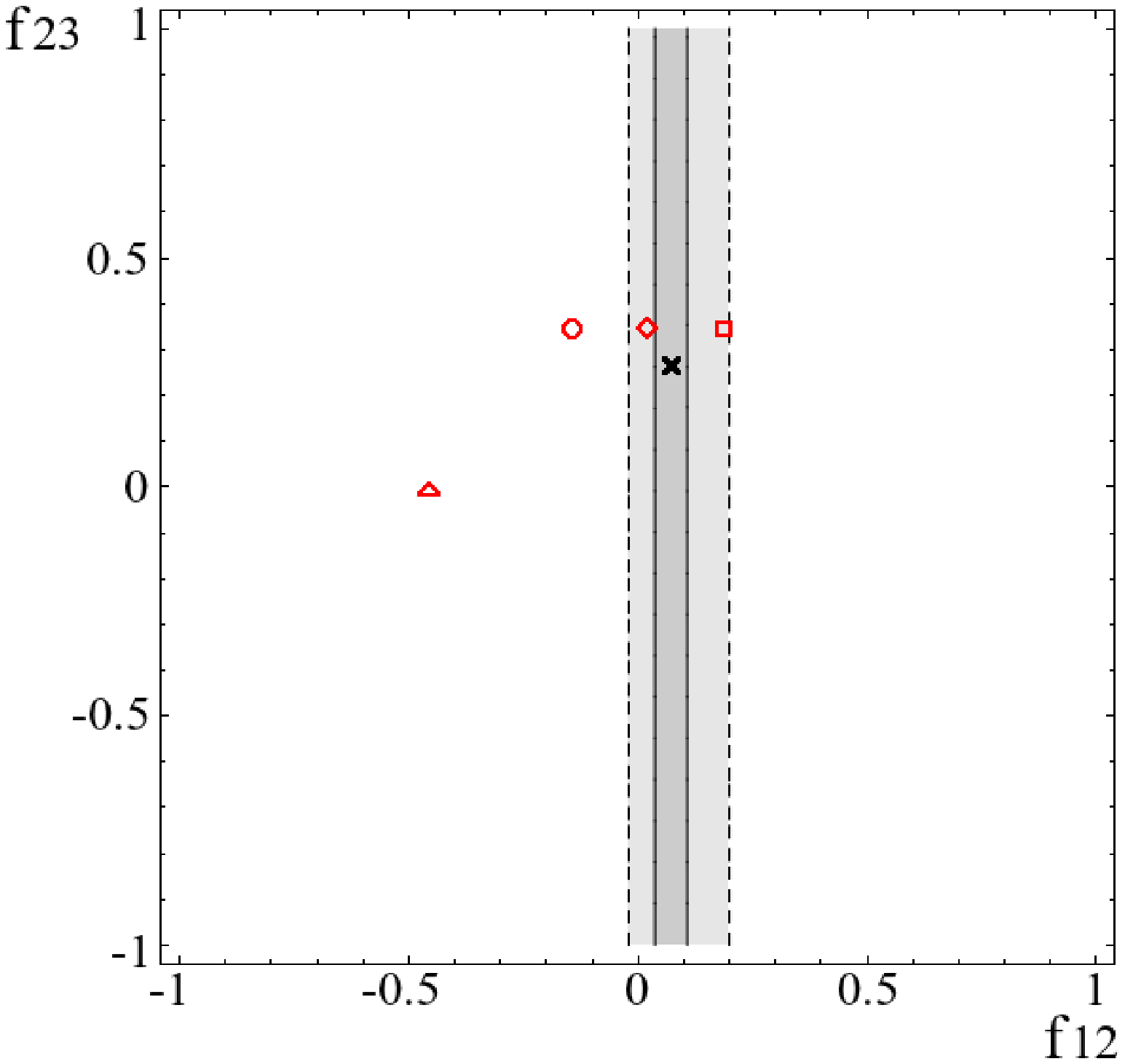}
	\caption{The fitted $1\sigma$ (solid line) and $3\sigma$ (dashed line) ranges for $f_{12}$ and $f_{23}$ of the standard oscillation with the measurement of $R_\pi$. The central value $(f_{12},~f_{23})$ is obtained from Eq. (\ref{stdosc}). Decay models in the normal hierarchy are tested in the left panel while those in the inverted hierarchy are tested in the right panel.}
	\label{fig:stdosc1}
	\end{center}
\end{figure}

To completely determine the relevant parameters and probe the flavor transition mechanisms, observations of two different sources of neutrinos are required. However, to accumulate enough data from two different sources may take a long time. As the pion source is the most common source, we first investigate the possibility of probing flavor transition models with the observation of neutrinos from the pion source alone.  In this case, the $\chi^2$-fitting formula is given by 
\begin{equation}
\chi^2=\Bigg(\frac{R_{\pi,\rm th}-R_{\pi,\rm exp}}{\sigma_{R_{\pi,\rm exp}}}\Bigg)^2,
\end{equation}
with the assumed accuracy $\sigma_{R_{\pi,\rm exp}}=(\Delta R_\pi/R_\pi)=10\%$ on the measurement $R_\pi$. For the input (true) flavor transition mechanism being the standard oscillation model, the fitting region of $f_{12}$ and $f_{23}$ is presented in Fig. \ref{fig:stdosc1}. Decay models in normal and inverted hierarchies are tested in the left and right panels respectively. Clearly, models in $dec1$ scenario are 
ruled out since the triangles are located far from the $3\sigma$ areas in both panels. Branching ratio $s$ representing models in $dec2$ is also constrained. From the left panel, it can be seen that $dec2$ scenario with the normal hierarchy can be more easily ruled out. However, models with the inverted hierarchy cannot be ruled out easily since no information on $f_{23}$ can be deduced from the measurement of $R_\pi$ alone.

\begin{figure}[htbp]
	\begin{center}
	\includegraphics[width=8cm]{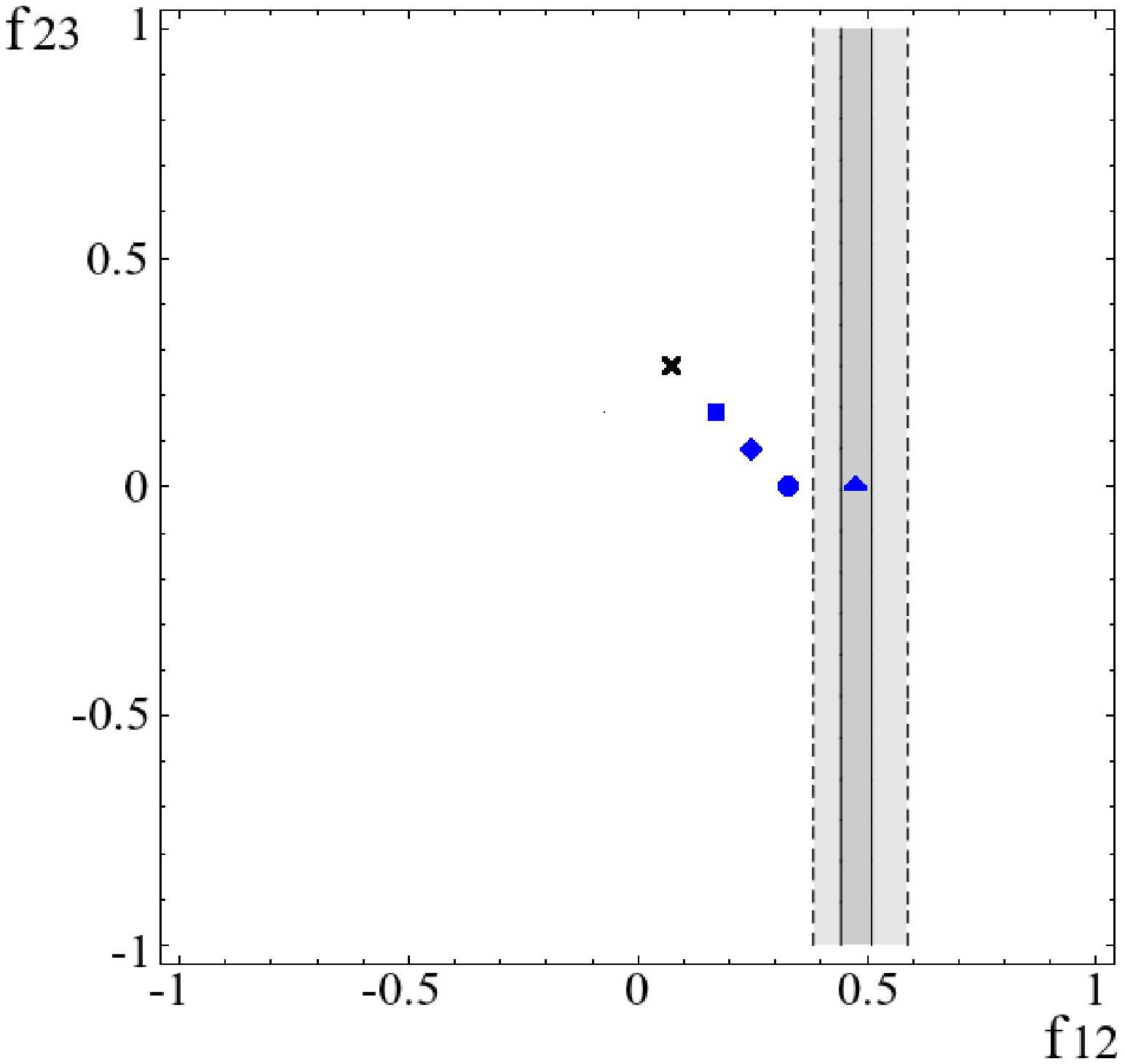}
	\includegraphics[width=8cm]{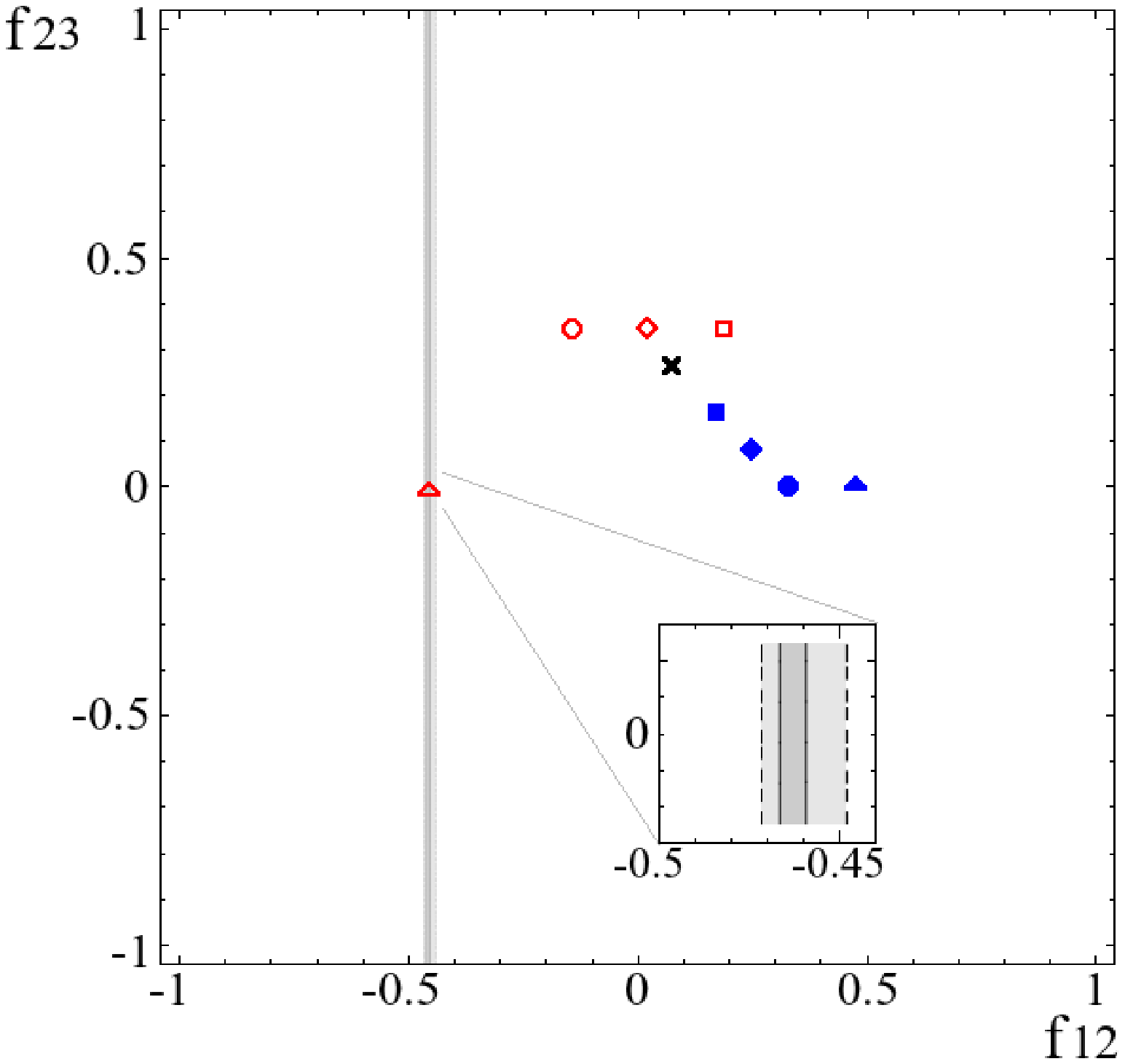}
	\caption{The fitted $1\sigma$ (solid line) and $3\sigma$ (dashed line) ranges for $f_{12}$ and $f_{23}$ for $dec1$ scenario with the measurement of $R_\pi$. The model described by Eq. (\ref{dec1norm}) in the normal hierarchy is the input model in the left panel while the model described by Eq. (\ref{dec1inv}) in the inverted hierarchy is the input model in the right panel. The inserted figure in the right panel provides a better resolution for the fitted area.}
	\label{fig:dec1to1}
	\end{center}
\end{figure}

Having considered the standard oscillation as the input model, let us now take $dec1$ scenario to be the input true model. 
 The fitting result is shown in Fig. \ref{fig:dec1to1}. The standard oscillation model and the models in the $dec2$ scenario are ruled out as $R_\pi$ measured with the assumed accuracy. 

\begin{figure}[htbp]
	\begin{center}
	\includegraphics[width=7.5cm]{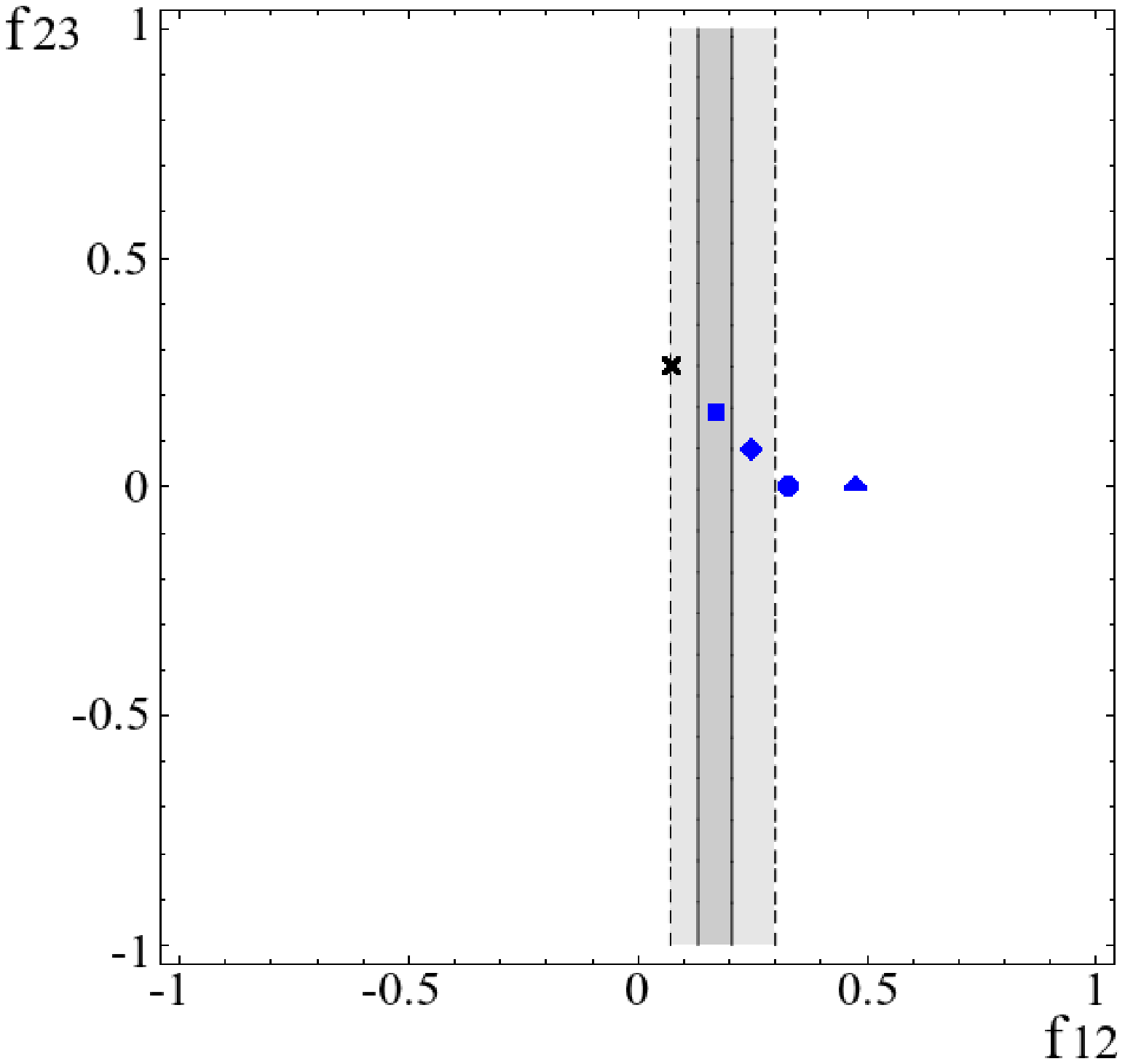}
	\includegraphics[width=7.5cm]{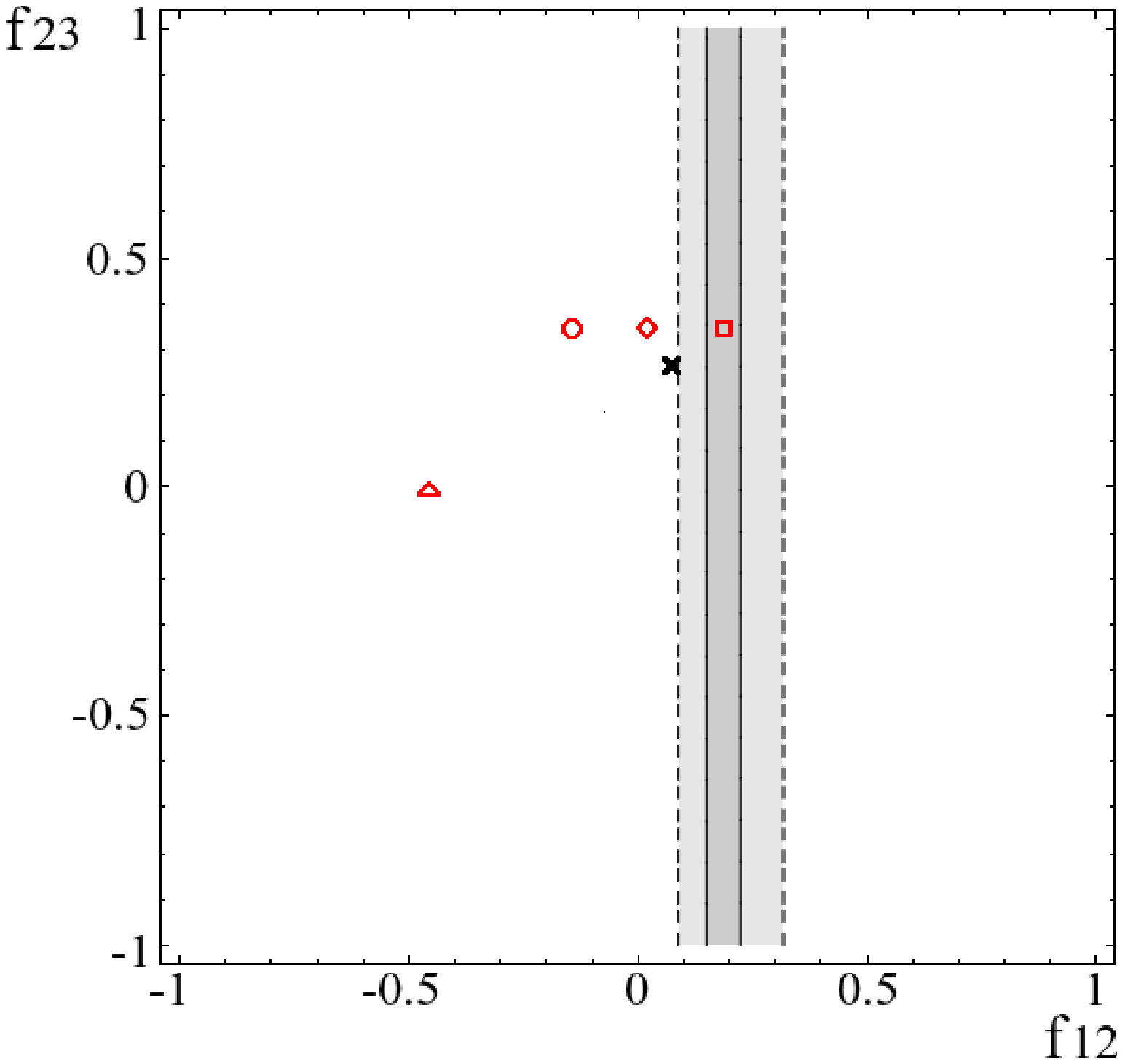}
	\includegraphics[width=7.5cm]{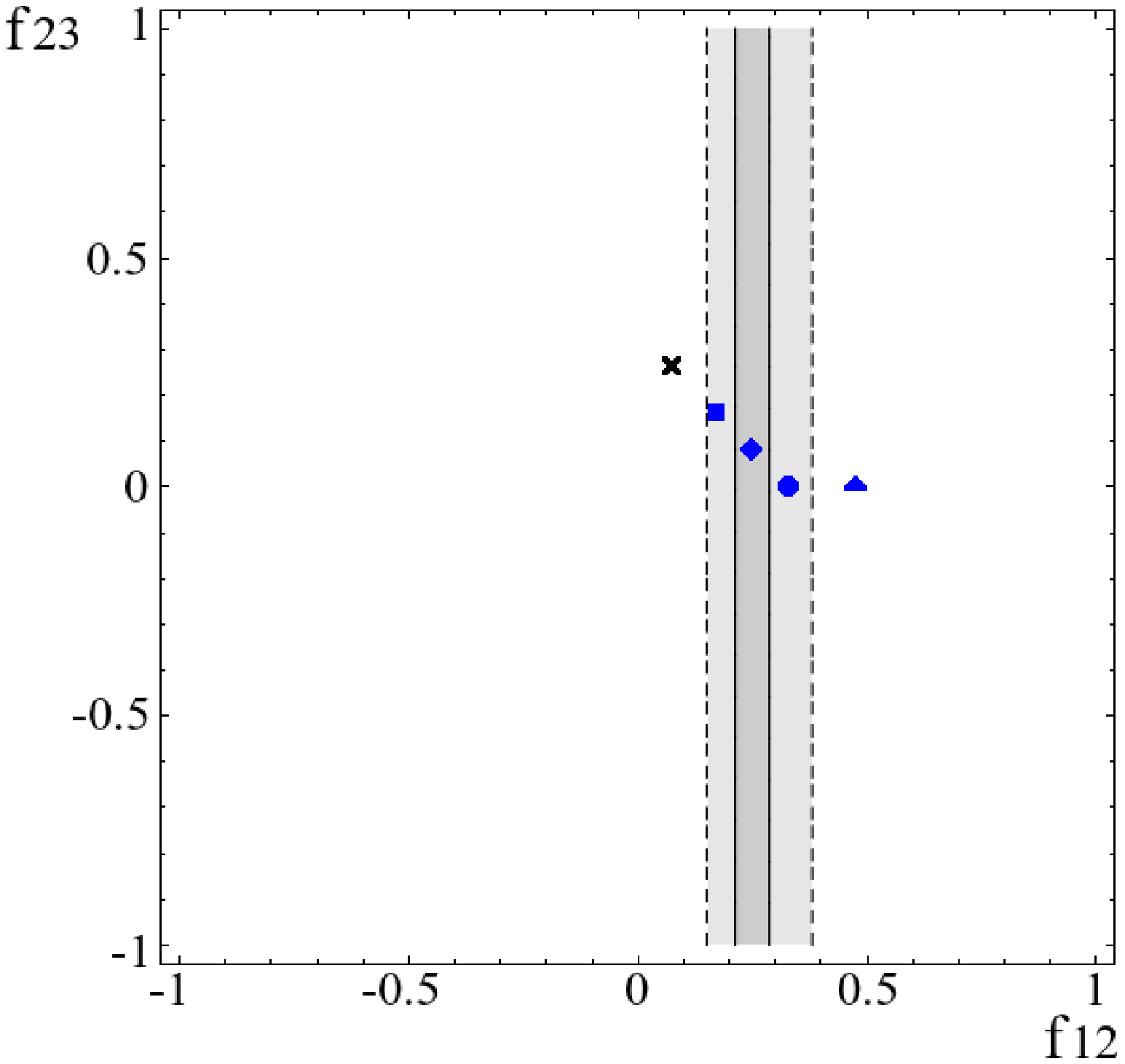}
	\includegraphics[width=7.5cm]{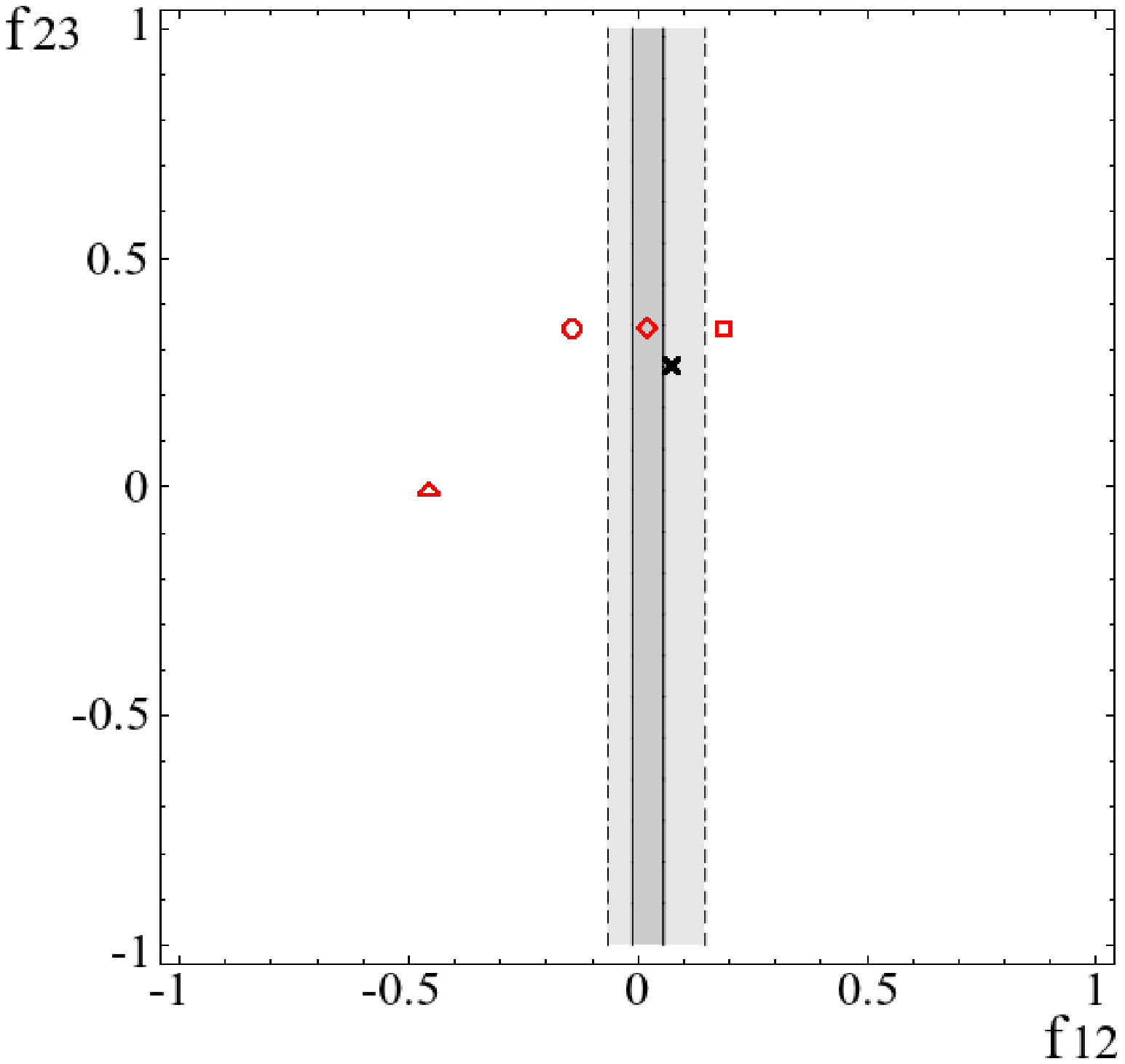}
	\includegraphics[width=7.5cm]{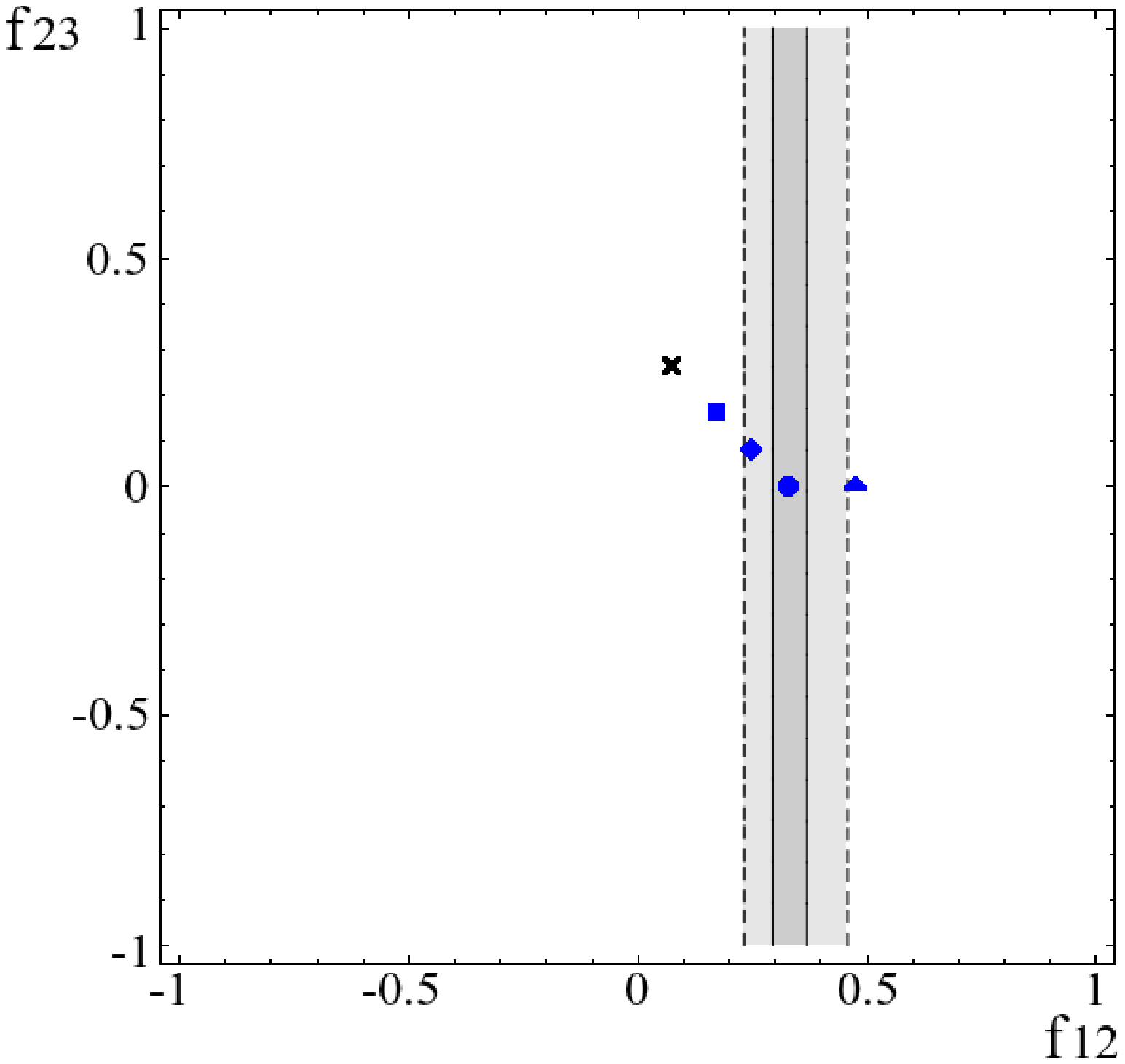}
	\includegraphics[width=7.5cm]{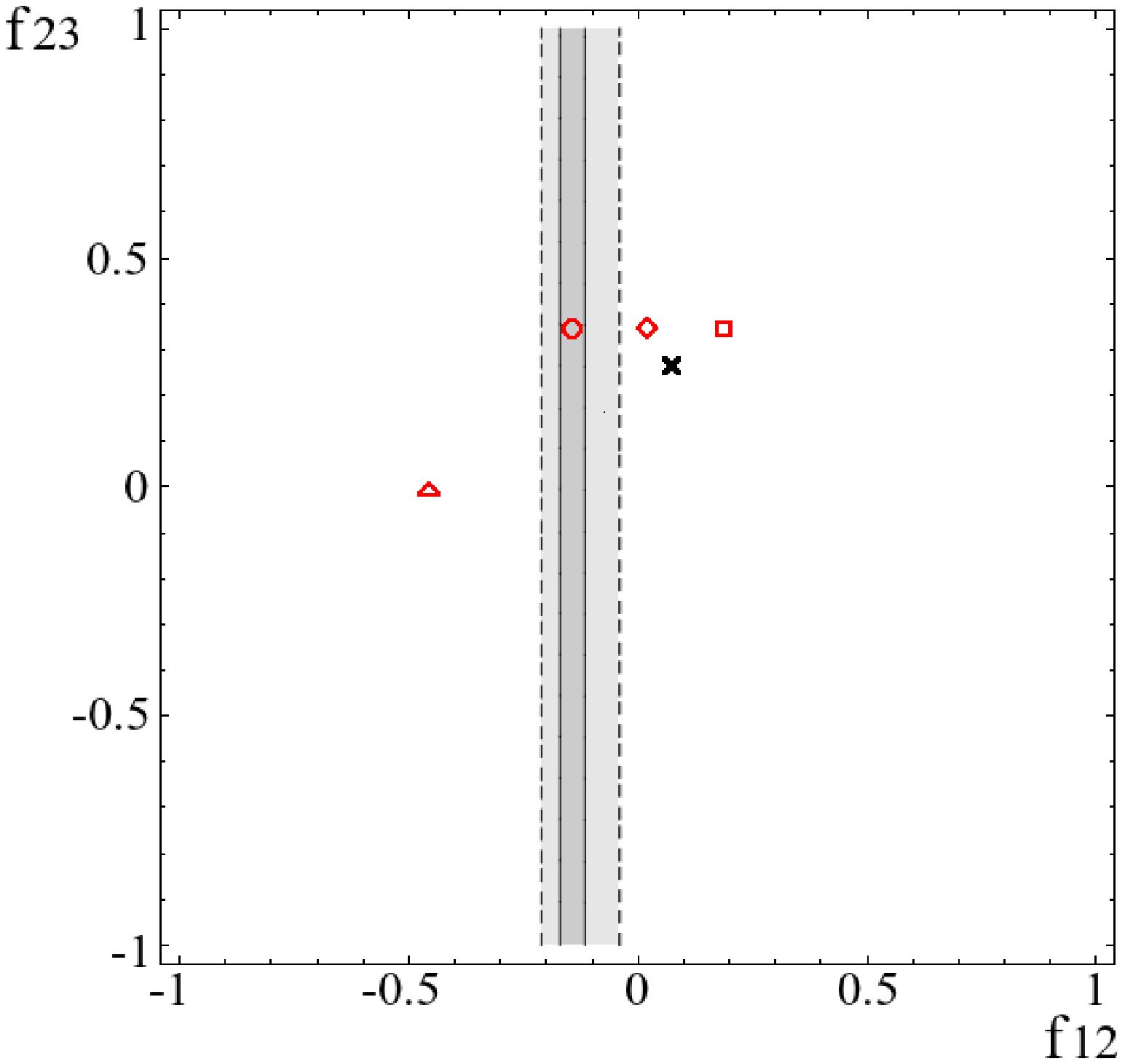}
	\caption{The fitted $1\sigma$ (solid line) and $3\sigma$ (dashed line) ranges for $f_{12}$ and $f_{23}$ for the $dec2$ scenario with the measurement of $R_\pi$. The models in the normal hierarchy described by Eq. (\ref{dec2norm}) with branching ratios of $s= 1, ~0.5~ {\rm and}~ 0$ are fitted in the left panel from top to bottom while the models in the inverted hierarchy described by Eq. (\ref{dec2inv}) with branching ratios of $s= 1, ~0.5~ {\rm and}~ 0$ are fitted in the right panel from top to bottom.}
	\label{fig:dec2to1}
	\end{center}
\end{figure}

Next, we take the input model to be one of those models in the $dec2$ scenario in which the heaviest state decays into two lighter ones with different branching ratios. The fitted regions for $f_{12}$ and $f_{23}$ are displayed in Fig. \ref{fig:dec2to1}. The left panel displays input models in the normal hierarchy and the right panel displays those in the inverted hierarchy. We sample, in both hierarchies, three different models with branching ratios of $s=1,~0.5~{\rm and}~0$ from top to bottom. On the left panel, we find that the standard oscillation can be ruled out for the input models with branching ratios of $s=0.5~{\rm and}~0$. However it cannot be ruled out for an input model with $s=1$. Furthermore the decay model described by Eq. (\ref{dec1norm}) is ruled out for input models with branching ratios of $s=0.5~{\rm and}~1$. However it cannot be ruled out for the input model with $s=0$. The upper plot on the left panel indicates that models with $s\lesssim0.5$ are ruled out for an input model with $s=1$. On the right panel, the scenario described by Eq. (\ref{dec2inv}) cannot be easily distinguished from the standard oscillation. In addition, the decay models in $dec1$ scenario described by Eq. (\ref{dec1norm}) and (\ref{dec1inv}) are also ruled out for input models in the $dec2$ scenario in both hierarchies.

\begin{figure}[htbp]
	\begin{center}
	\includegraphics[width=8cm]{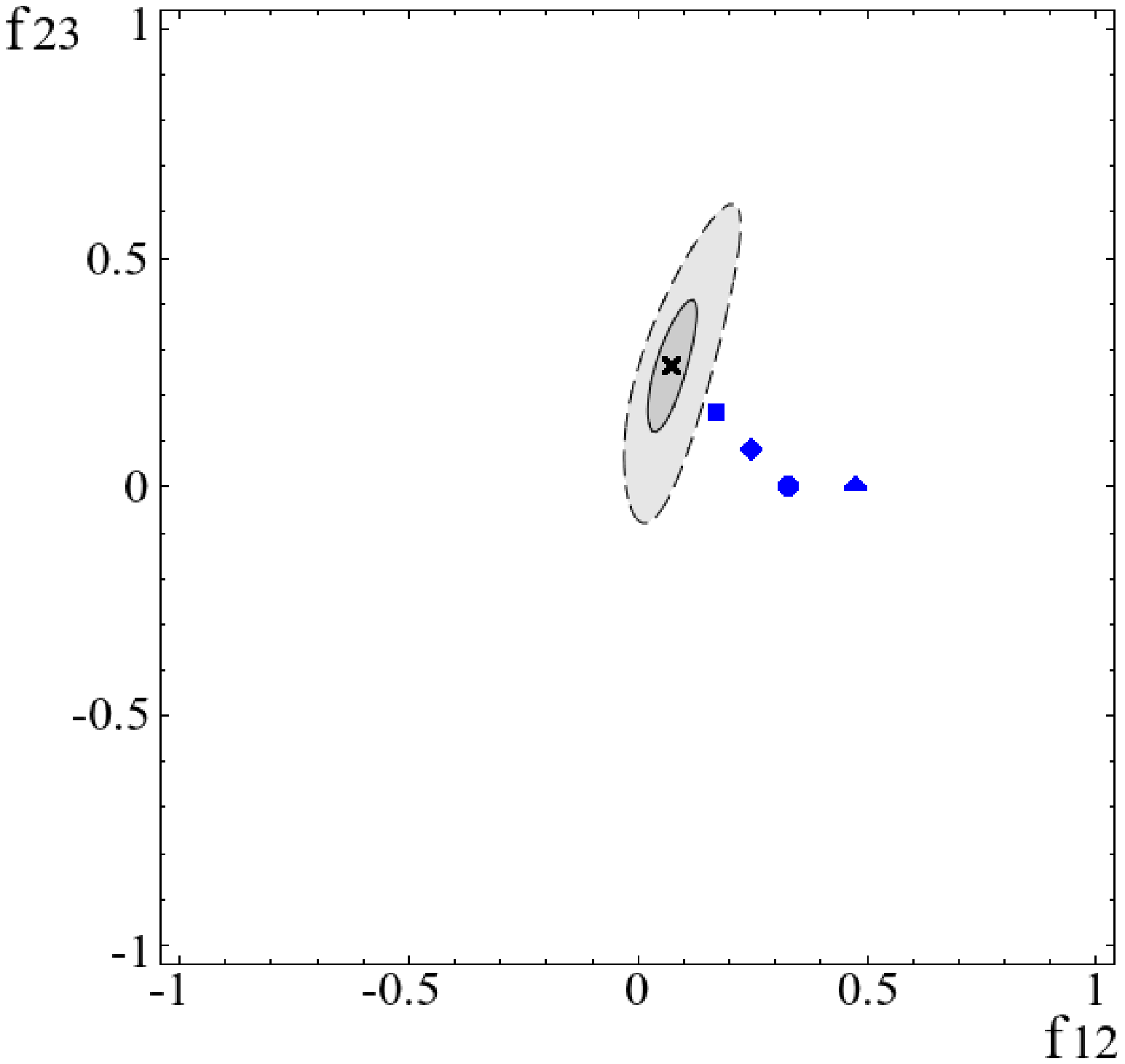}
	\includegraphics[width=8cm]{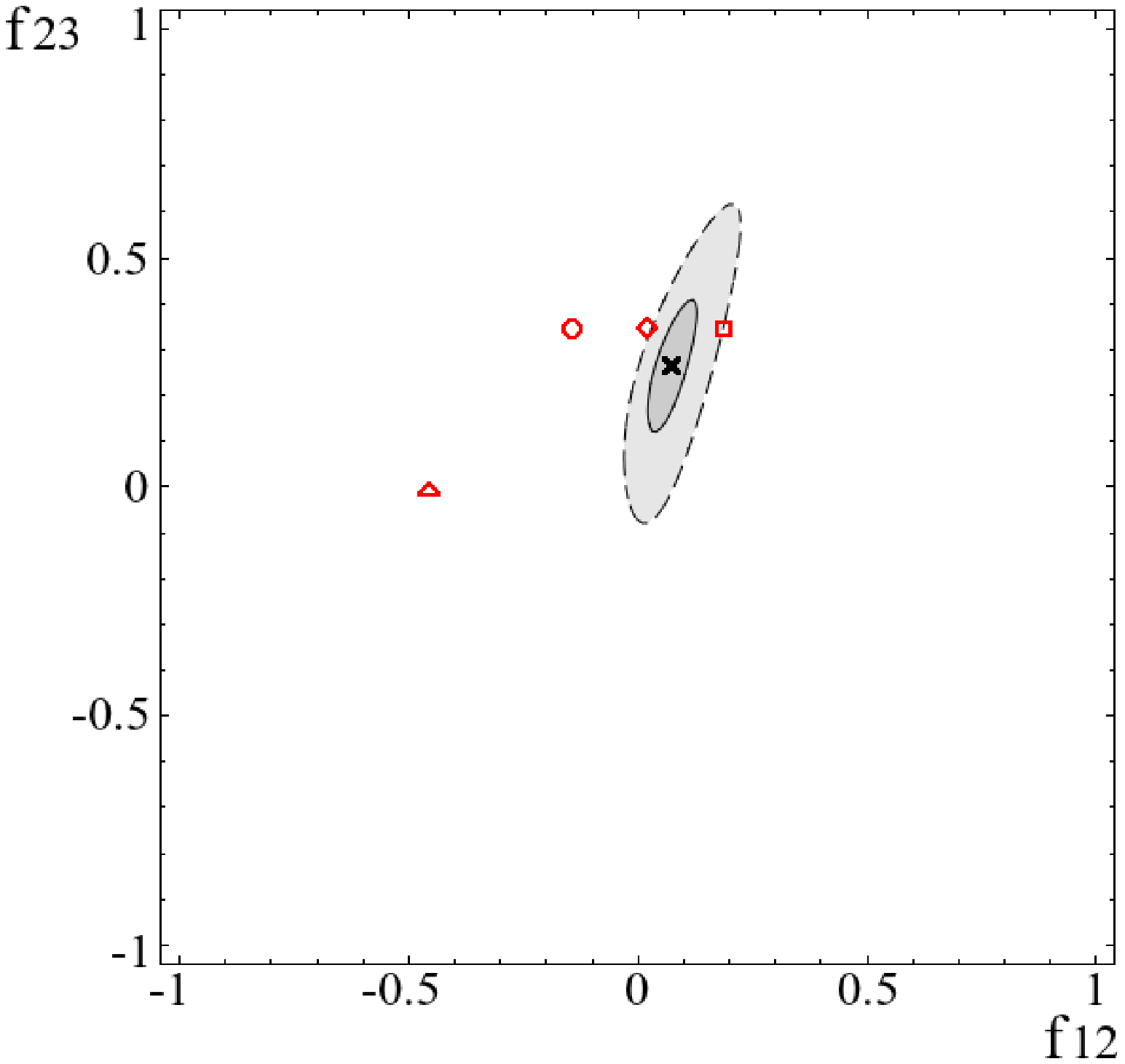}
	\caption{The fitted $1\sigma$ (solid line) and $3\sigma$ (dashed line) ranges for $f_{12}$ and $f_{23}$ of the standard oscillation with the measurements of both $R_\pi$ and $R_\mu$. The central value $(f_{12},~f_{23})$ is obtained from Eq. (\ref{stdosc}). Decay models in the normal hierarchy are tested in the left panel while those in the inverted hierarchy are tested in the right panel.}
	\label{fig:stdosc2}
	\end{center}
\end{figure}

We next consider the observation from both the pion source and the muon damped source \cite{Kashti:2005qa}. The determinations of both $R_\pi$ and $R_\mu$ consequently determine $f_{12}$ and $f_{23}$ completely via Eq. (\ref{Rtof}). Applying the $\chi^2$-analysis, Eq. (\ref{chi2}), with assumed accuracies $\sigma_{R_{\pi,\rm exp}}=(\Delta R_\pi/R_\pi)R_{\pi,\rm exp}=\sigma_{R_{\mu,\rm exp}}=(\Delta R_\mu/R_\mu)R_{\mu,\rm exp}=10\%$, we probe the flavor transition models again with simultaneous measurements of $R_\pi$ and $R_\mu$. One dose not need to include uncertainties of mixing angles $\theta_{ij}$ and $CP$ phase $\delta$ since their effects are already embedded in $f_{12}$ and $f_{23}$. Let us begin by taking standard neutrino oscillation as the input model. The fitted region for $f_{12}$ and $f_{23}$ is presented in Fig. \ref{fig:stdosc2}. Under this measurement accuracy, the decay scenarios can be ruled out at $3\sigma$ level for the normal hierarchy as shown in the left panel. But, for the inverted hierarchy,  only the model described by Eq. (\ref{dec1inv}) can be ruled out.

\begin{figure}[htbp]
	\begin{center}
	\includegraphics[width=8cm]{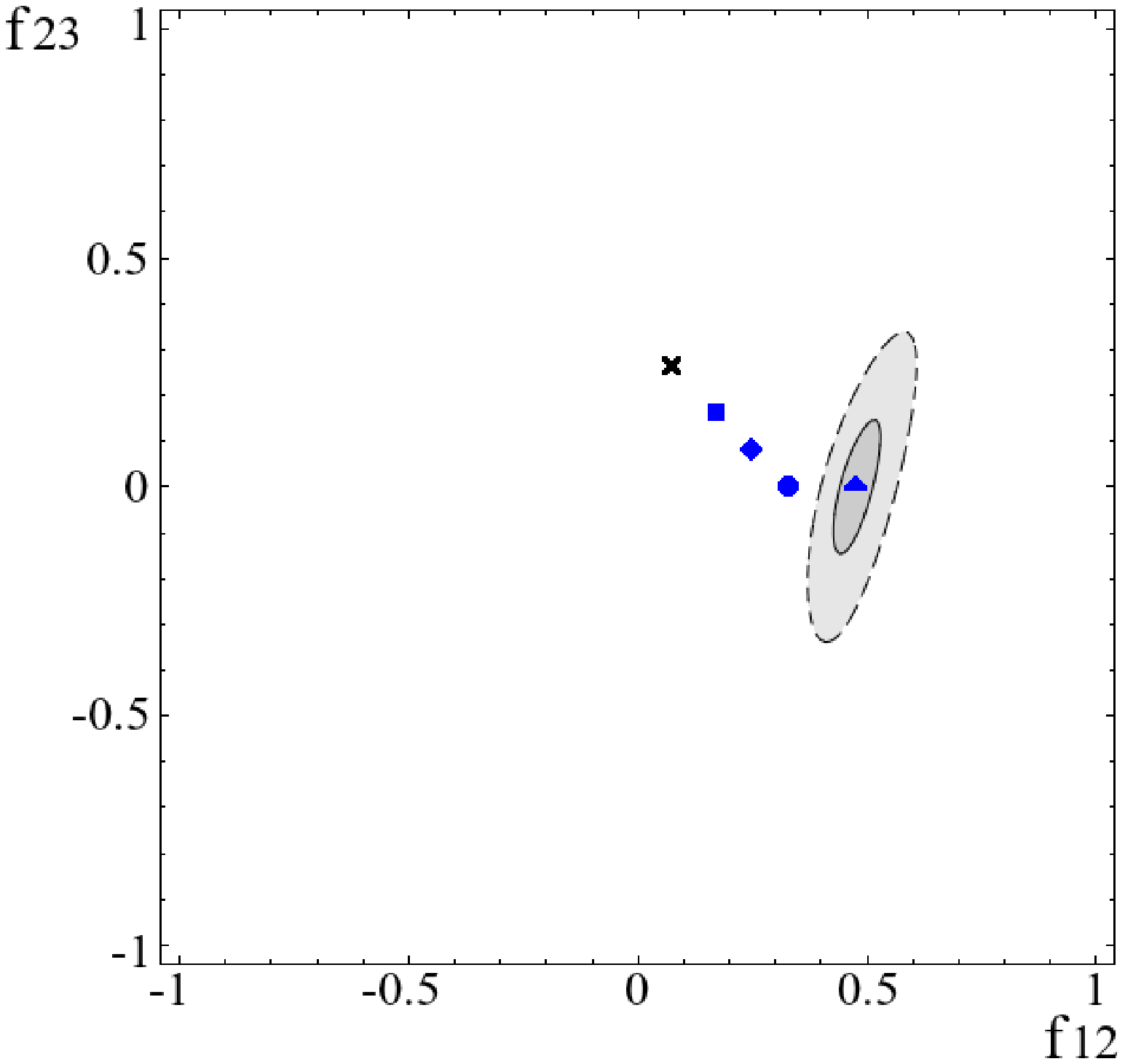}
	\includegraphics[width=8cm]{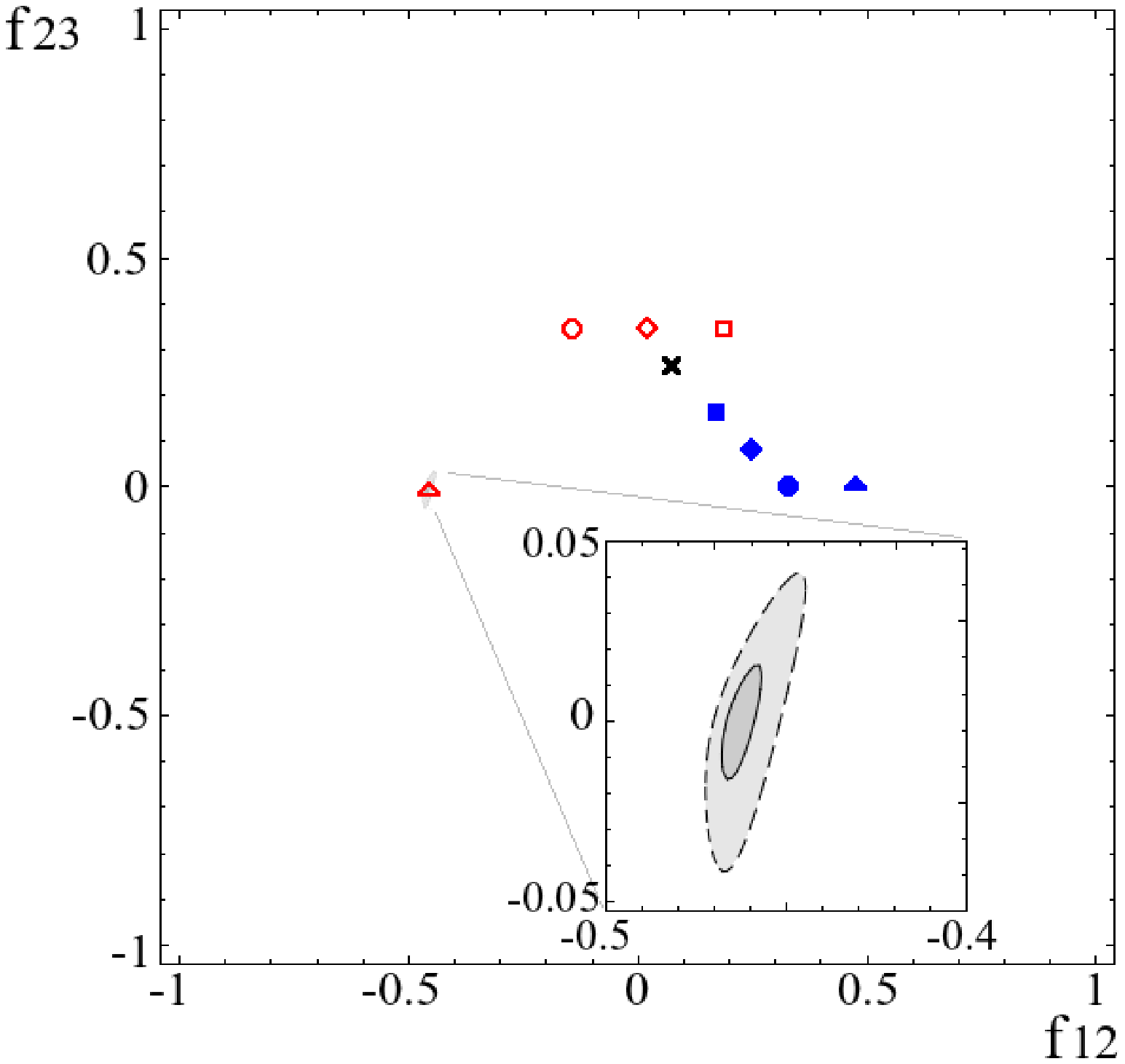}
	\caption{The fitted $1\sigma$ (solid line) and $3\sigma$ (dashed line) ranges for $f_{12}$ and $f_{23}$ for $dec1$ scenario with the measurements of both $R_\pi$ and $R_\mu$. The input model in the left panel is in the normal hierarchy described by Eq. (\ref{dec1norm}) while the input model in the right panel is in the inverted hierarchy described by Eq. (\ref{dec1inv}). The inserted figure in the right panel provides a better resolution for the fitted area.}
	\label{fig:dec1to2}
	\end{center}
\end{figure}

We next take the input models as those in $dec1$ scenario. In the left panel of Fig. \ref{fig:dec1to2}, the input model is that described by Eq. (\ref{dec1norm}) while the input  model in the right panel is
that described by Eq. (\ref{dec1inv}). Both indicate that, with $10\%$ uncertainties in measurement, standard oscillation and the models in $dec2$ scenario can be ruled out.


\begin{figure}[htbp]
	\begin{center}
	\includegraphics[width=7.5cm]{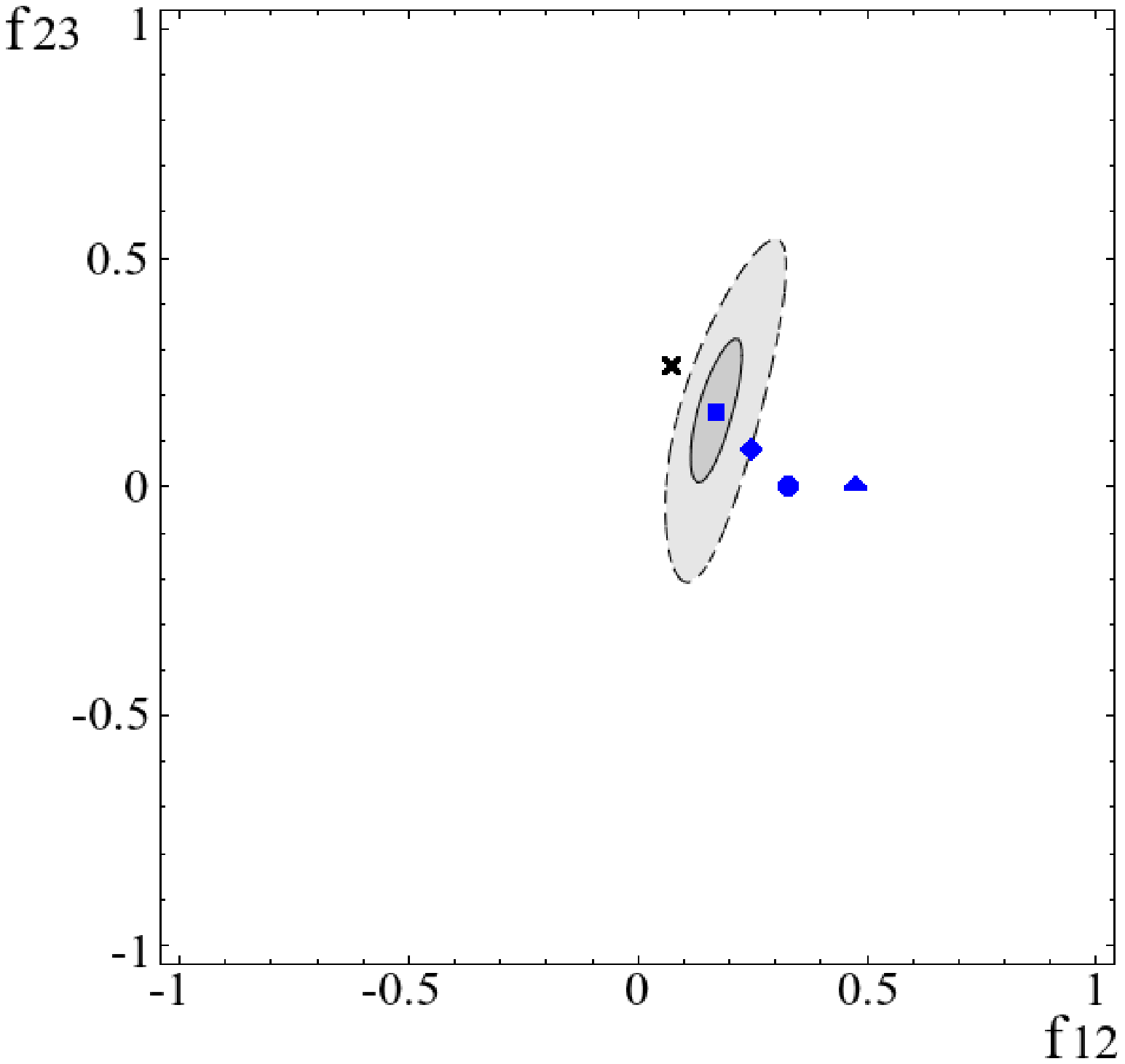}
	\includegraphics[width=7.5cm]{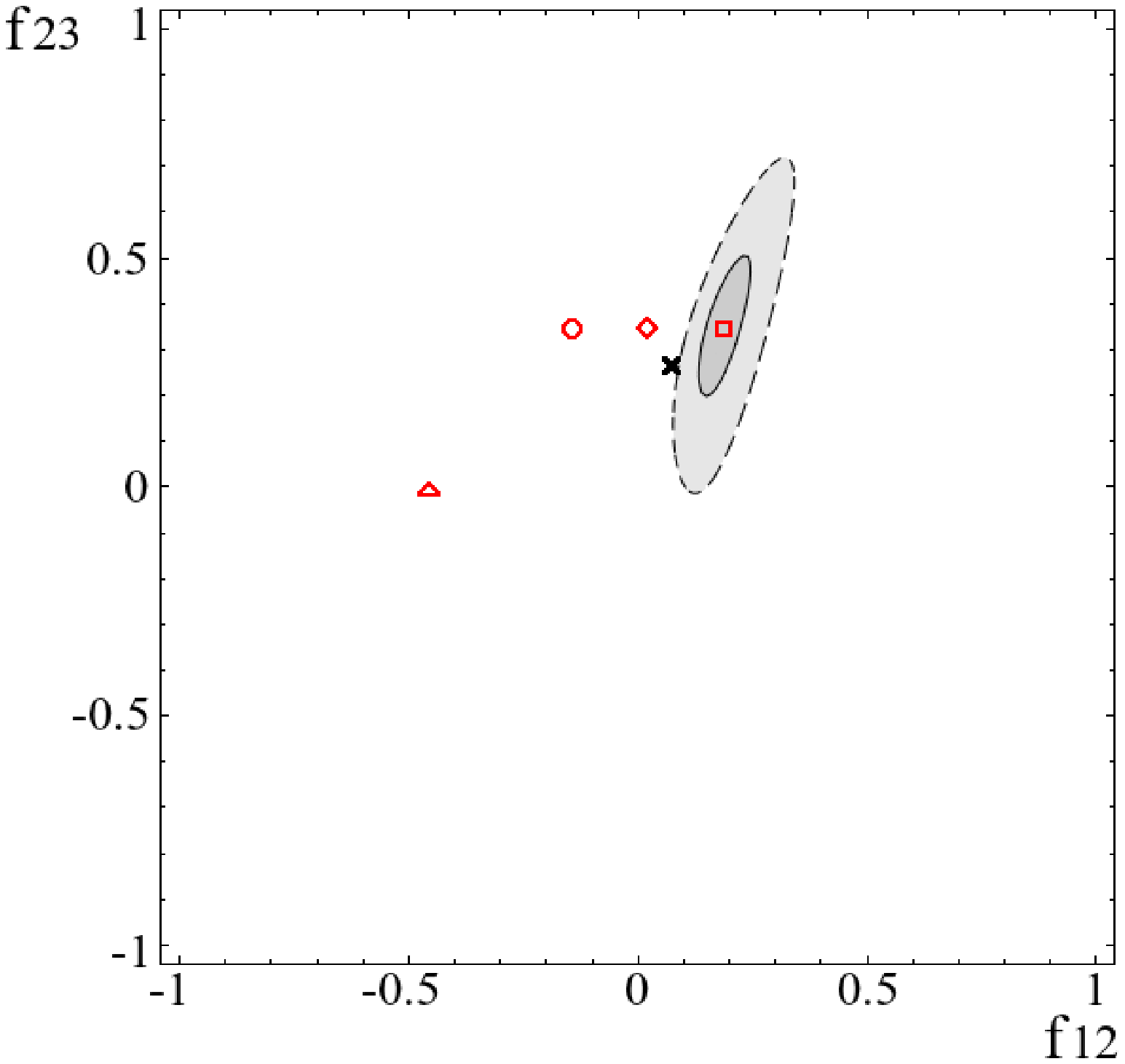}
	\includegraphics[width=7.5cm]{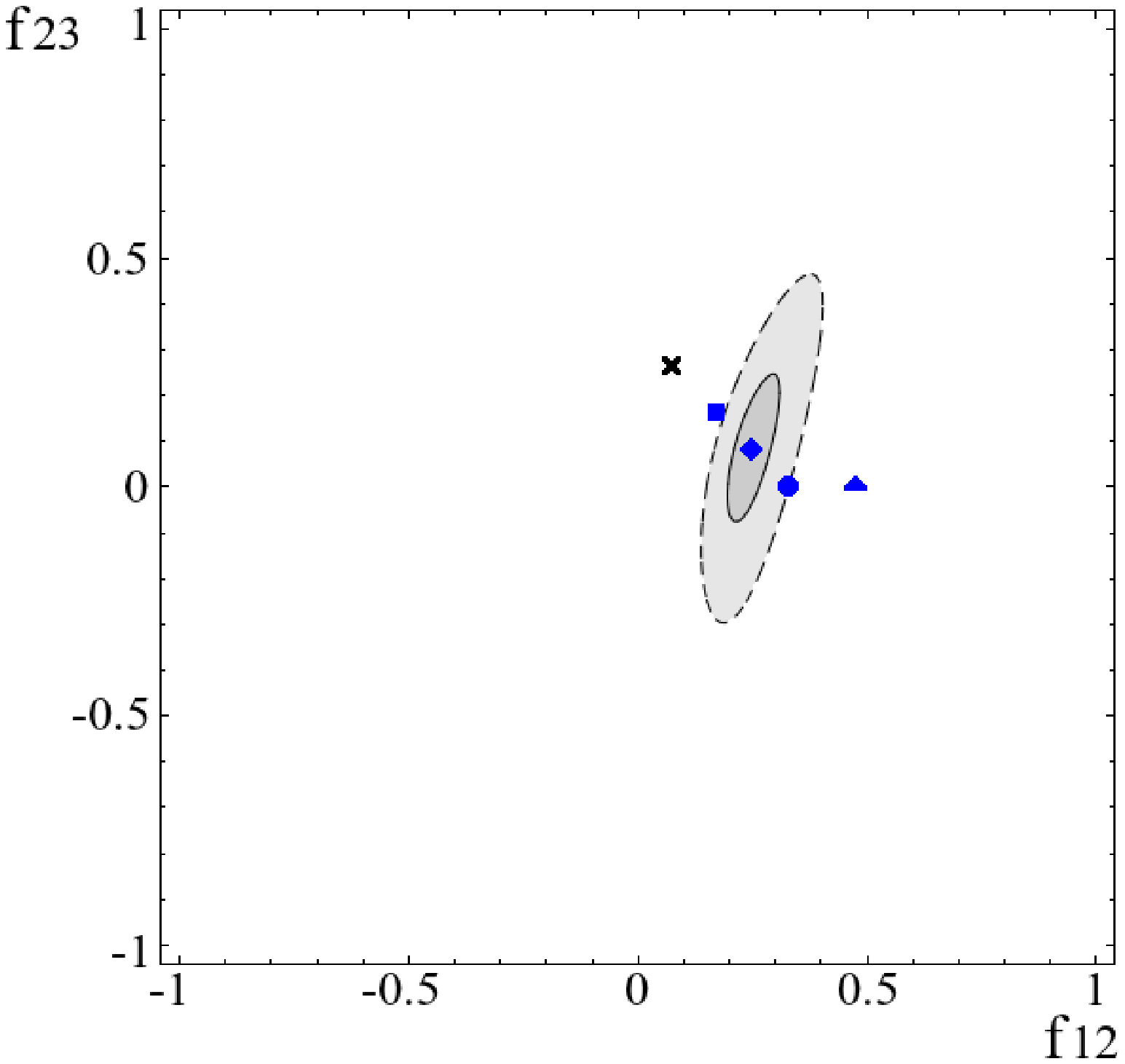}
	\includegraphics[width=7.5cm]{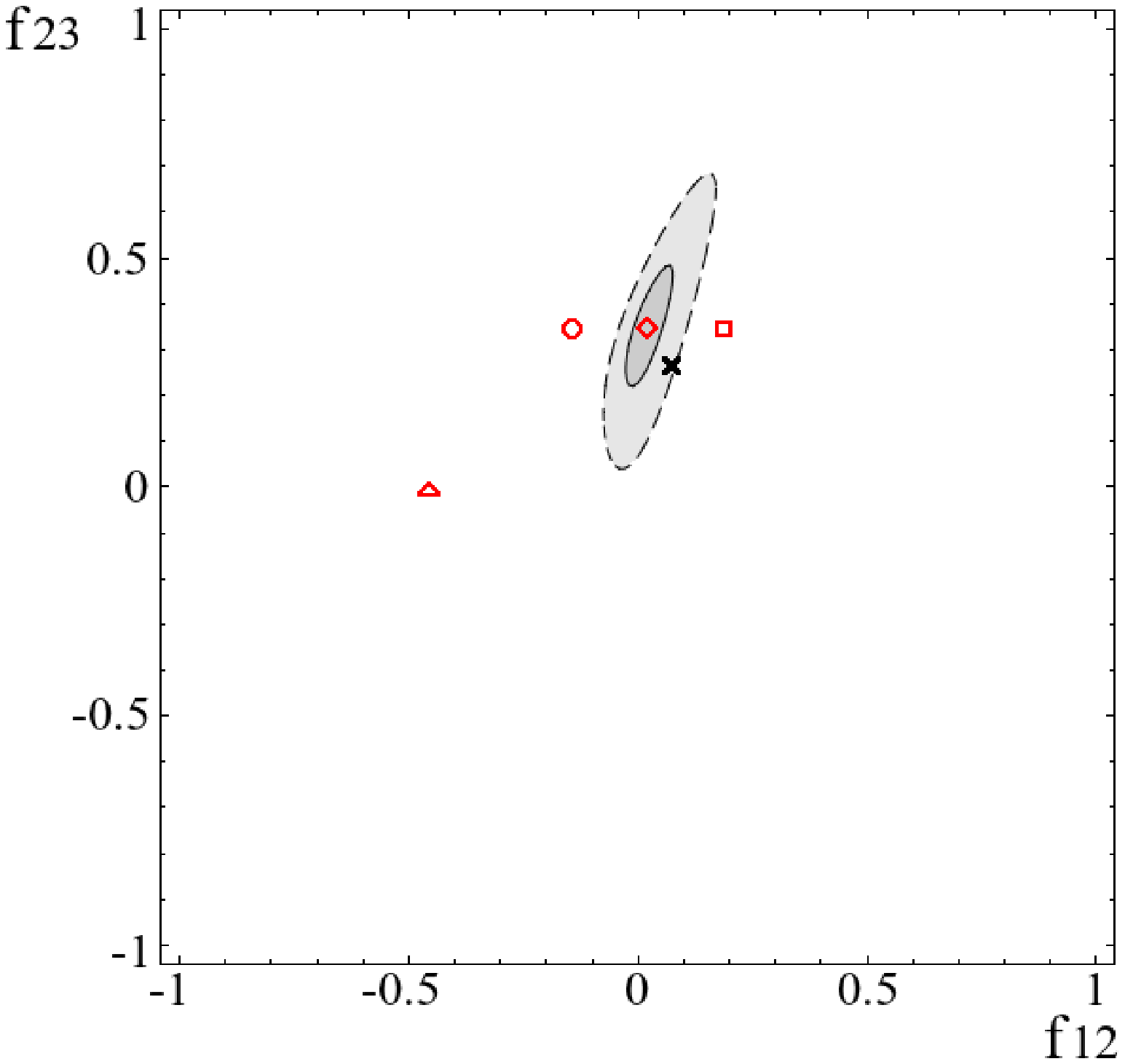}
	\includegraphics[width=7.5cm]{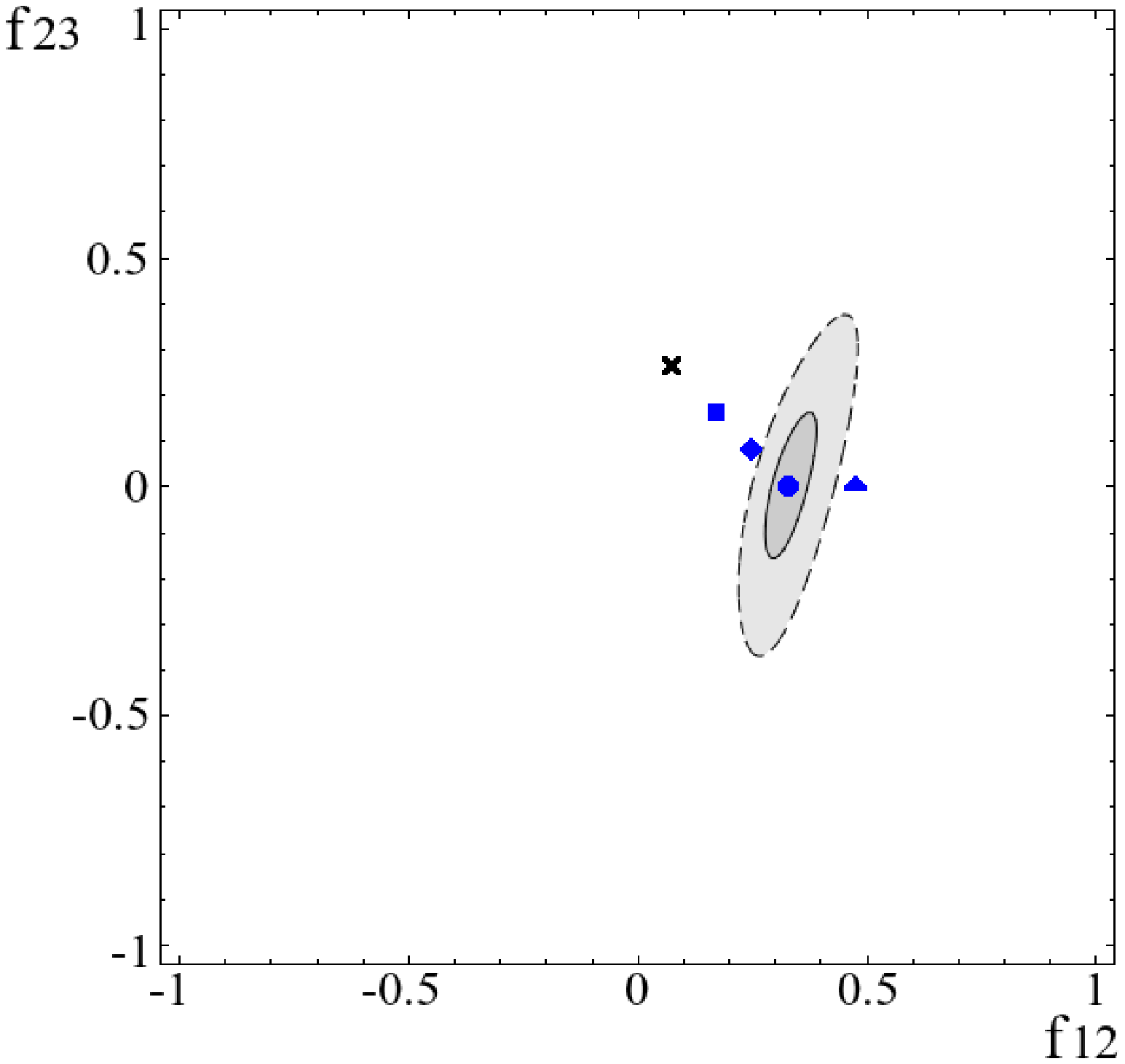}
	\includegraphics[width=7.5cm]{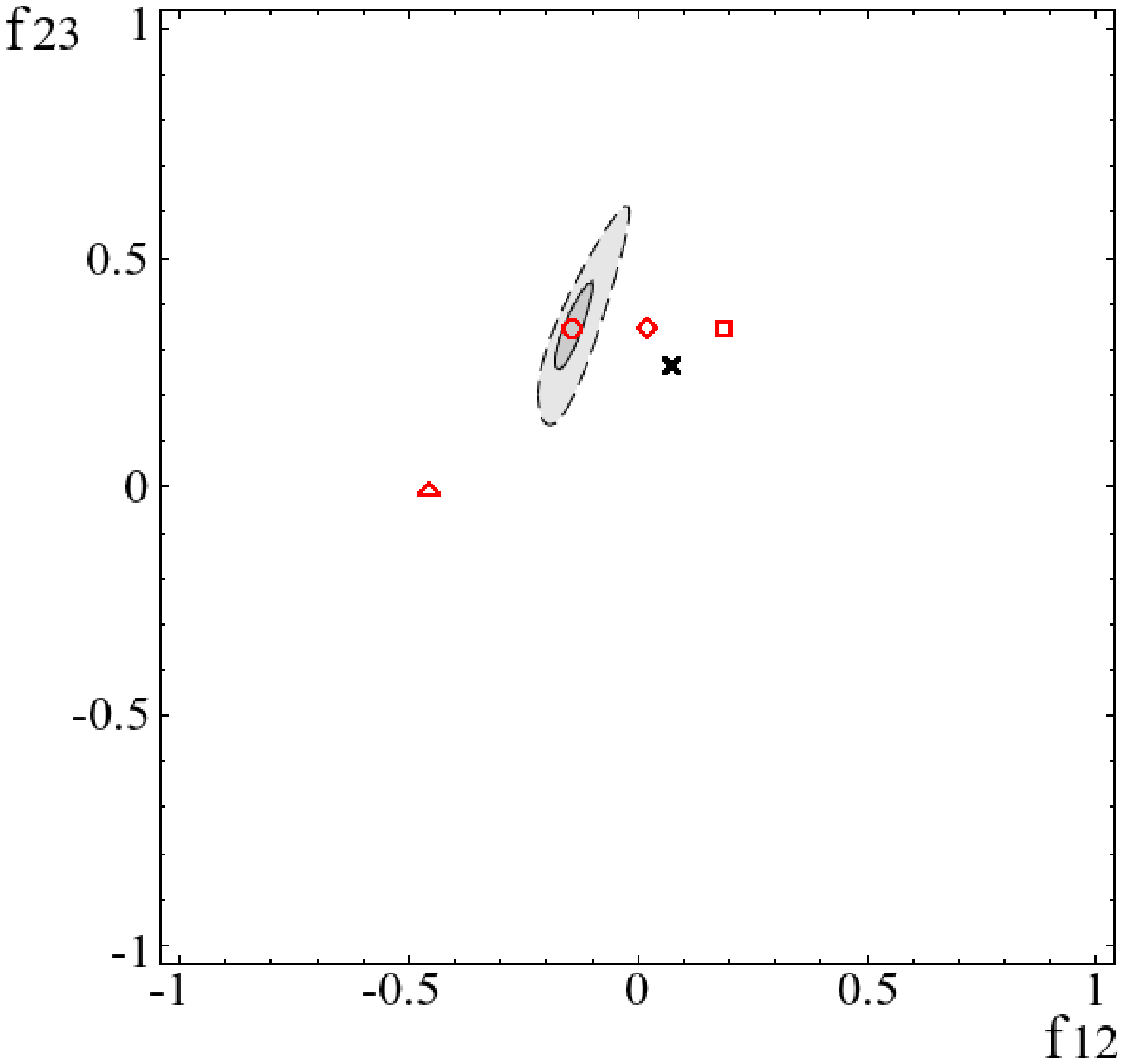}
	\caption{The fitted $1\sigma$ (solid line) and $3\sigma$ (dashed line) ranges for $f_{12}$ and $f_{23}$ for the $dec2$ scenario with both $R_\pi$ and $R_\mu$. From top to bottom in the left panel, the models in the normal hierarchy described by Eq. (\ref{dec2norm}) with branching ratios of $s= 1, ~0.5~ {\rm and}~ 0$ are taken as input  models. The input models in the right panel are in the inverted mass hierarchy described by Eq. (\ref{dec2inv}) with branching ratios of $s= 1, ~0.5~ {\rm and}~ 0$  from top to bottom.}
	\label{fig:dec2to2}
	\end{center}
\end{figure}

In Fig. \ref{fig:dec2to2}, the fitted region for $f_{12}$ and $f_{23}$ are presented for $dec2$ scenario in which the heaviest mass state decays into two lighter ones with different branching ratios. The left panel displays input models in the normal hierarchy and the right panel displays those in the inverted hierarchy. We sample, in both hierarchies, three different models with branching ratios of $s=1,~0.5~{\rm and}~0$ from top to bottom. From the left panel, we find that, in the normal hierarchy, the standard oscillation and the decay model described by Eq. (\ref{dec1norm}) are ruled out for input models in the $dec2$ scenario described by Eq. (\ref{dec2norm}). The decay branching ratio in scenario described by Eq. (\ref{dec2norm}) are also constrained. For example, the upper plot on the left panel indicates that models with $s\lesssim0.5$ are ruled out in this scenario for an input model with $s=1$. For the inverted hierarchy in the right panel, the standard oscillation cannot be ruled out for input models in the $dec2$ scenario described by Eq. (\ref{dec2inv}). Only the decay model described by Eq. (\ref{dec1inv}) is ruled out.

For normal mass hierarchy, we conclude that, from Fig. \ref{fig:stdosc2}--\ref{fig:dec2to2}, the standard oscillation, the $dec1$ and $dec2$ scenarios can be discriminated between one another with the assumed accuracy of measurement. For inverted mass hierarchy, only the $dec1$ scenario can be distinguished from the standard oscillation and the $dec2$ scenario while the latter two cannot be discriminated from each other.

\section{Summary and Conclusions}

In summary, we have illustrated that, for ultrahigh energy neutrinos, the neutrino flavor transition mechanism can be probed by measuring the $\nu_e$ fraction of the total neutrino flux. We parameterize the flavor transitions of propagating astrophysical neutrinos by the matrix $Q$, proposed in[26]. In the $Q$-representation, flavor transition models are classified by the third row of the $Q$ matrix. 

In the limit of exact $\nu_\mu-\nu_\tau$ symmetry, which we have adopted in the earlier work~\cite{Lai:2010tj}, one has $Q_{32}=0$ so that the relevant matrix elements for classifying flavor transition models are $Q_{31}$ and $Q_{33}$. In this paper, we generalize our earlier approach to the case without $\nu_\mu-\nu_\tau$ symmetry. We argue that the $\nu_e$ fraction of the total ultrahigh energy neutrino flux can be extracted by detecting shower-induced Cherenkov radiation in radio-wave neutrino telescopes. The new observable, $R\equiv\phi_e/(\phi_\nu+\phi_\tau)$, is introduced  for flavor discrimination in these radio-wave neutrino telescopes. We then argue that this flux ratio is directly related to parameters
$f_{12}\equiv Q_{31}-Q_{32}$ and $f_{23}\equiv Q_{32}+Q_{33}$.  It has been shown in 
Fig. \ref{fig:Cvalue} that flavor transition models are well classified by $(f_{12},~f_{23})$. It is clear that $f_{12}$ and $f_{23}$ reduce to $Q_{31}$ and $Q_{33}$, respectively, in the limit of exact 
$\nu_\mu-\nu_\tau$ symmetry. By measuring $R$, the new proposed observable for ultrahigh energy astrophysical neutrinos, we are able to probe neutrino flavor transition models classfied by new parameters $(f_{12},~f_{23})$ in a model independent fashion.
  
To test further the capability of discriminating between different flavor transition models, we fit $(f_{12},~f_{23})$ to measured ratios $R_{\pi,{\rm exp}}$ and $R_{\mu,{\rm exp}}$ using Eq. (23) for a few illustrative models. The ranges for $(f_{12},~f_{23})$ are presented up to the $3\sigma$ confidence level for three classes of input models. We first consider the case that only the pion source is measured. We then discuss the case that both the pion source and the muon damped source are measured. We have found that the measurement accuracy of $(\Delta R_\pi/R_\pi)=10\%$ is sufficient to discriminate $dec1$ scenario described by Eq. \ref{dec1norm} and \ref{dec1inv} from the standard neutrino oscillation and $dec2$ scenario given by Eq. (\ref{dec2norm}) and (\ref{dec2inv}). Though the accuracies are not sufficient to entirely discriminate among the standard oscillation model and models in the $dec2$ scenario, models within the $dec2$ scenario can be distinguished from one another. By including the measurement of muon damped source with the same accuracy, $(\Delta R_\mu/R_\mu)=10\%$,  the whole parameter space $(f_{12},~f_{23})$ is constrained. However,  the effectiveness for discriminating flavor transition models is not much improved. This implies that either an observation of the third source or a more accurate measurement is required to further distinguish between standard neutrino oscillation and other flavor transition models.

For cosmogenic neutrinos, one has $E^2_\nu dN_\nu/dE_\nu\simeq (10^{-8} -10^{-9})~{\rm GeV}~{\rm cm}^{-2}{\rm s}^{-1}{\rm sr}^{-1}$ for $E_\nu\simeq10^{18}{\rm eV}$ \cite{Ahlers:2012rz}. With ARA 3-year exposure, the projected number of neutrino events is around 50 for baseline models or around 150 for strong evolution models. Either case indicates that the accuracy $(\Delta R/R)=10\%$ is reachable in a decade of  ARA data taking\cite{Allison:2011wk}.

We like to point out that the observation of astrophysical neutrinos from non-$\nu_\tau$ sources is never sufficient to completely determine the third row of the $Q$ matrix. To determine each of $Q_{31}$, $Q_{32}$ and $Q_{33}$, the observation from at least one source with nonzero fraction of $\nu_\tau$ is required, in addition to two non-$\nu_\tau$sources. Furthermore, the statistical analysis outlined by Eq. (\ref{chi2}) is performed with a precise knowledge of the neutrino flavor ratio at the source. To take into account the uncertainty of the neutrino flavor ratio at the source, the statistical analysis should be refined. We shall address these issues in a future publication.

\section*{Acknowledgements}
This work is supported by the National Science Council of Taiwan under Grants No. NSC-100-18-2112-M-182-001-MY3, NSC-99-2112-M-009-005-MY3, and Focus Group on Cosmology and Particle Astrophysics, National Center for Theoretical Sciences, Taiwan.

\end{document}